\documentclass[useAMS,usenatbib,usegraphicx]{mn2e} 

\usepackage{mathptmx}
\usepackage{amssymb}
\usepackage{color}
\bibpunct{(}{)}{;}{a}{}{,}
\newcommand{\aap}{A\&A}

\newcommand{\mnras}{MNRAS}

\newcommand{\apss}{Ap\&SS}
\newcommand{\apj}{ApJ}
\newcommand{\araa}{ARA\&A}

\newcommand{\aj}{AJ}

\newcommand{\apjl}{ApJ}

\newcommand{\ignore}[1]{}

\newcommand{\rev}[1]{{#1}}
\newcommand{\rvtwo}[1]{{ #1}}
\newcommand{\rem}[1]{{}}

\title[Galactic winds - How to launch galactic outflows in typical Lyman-break galaxies]{Galactic winds - How to launch galactic outflows in typical Lyman-break galaxies}
\author[W. von Glasow, M. G. H. Krause, J. Sommer-Larsen, A. Burkert]{W. von Glasow$^{1,2}$\thanks{E-mail:
glasow@usm.uni-muenchen.de}, M. G. H. Krause$^{1,3,2}$\thanks{E-mail: krause@mpe.mpg.de}, J. Sommer-Larsen$^{2,4}$\thanks{E-mail: j.sommerlarsen@gmail.com}, A. Burkert$^{1,2,3,5}$\thanks{E-mail: burkert@usm.uni-muenchen.de}\\
$^{1}$Universit{\"a}ts-Sternwarte, Ludwig-Maximilians-Universit{\"a}t (LMU), Scheinerstrasse 1, 81679 M{\"u}nchen, Germany\\
$^{2}$Excellence Cluster Universe, Technische Universit{\"a}t M{\"u}nchen, Boltzmannstrasse 2, 85748 Garching, Germany\\
$^{3}$Max-Planck-Institut f{\"u}r extraterrestrische Physik, Giessenbachstrasse, 85748 Garching, Germany\\
$^{4}$Dark Cosmology Centre, Niels Bohr Institute, University of Copenhagen, Juliane Maries Vej 30, DK-2100 Copenhagen, Denmark\\
$^{5}$Max-Planck-Fellow}
\begin{document}

\date{June 20th, 2012, rev1: April 29th, 2013}

\pagerange{\pageref{firstpage}--\pageref{lastpage}} \pubyear{2012}

\maketitle

\label{firstpage}

\begin{abstract}

We perform hydrodynamical simulations of a young galactic disc embedded in a hot gaseous halo
\rev{using parameters typical for Lyman break galaxies (LBGs).}
We take into account the (static) gravitational potentials due to a dark matter halo, a stellar bulge and a disc of stars and gas. 
Star formation is treated by a local Kennicutt-Schmidt law. 
\rvtwo{We simplify the structure of the interstellar medium by restricting the computational domain to a 25th of the full azimuthal angle, effectively assuming large-scale axisymmetry and neglecting any effects of spiral structure, and focus on the large-scale ISM drivers, the superbubbles.}
Supernovae are triggered randomly and have preset event sizes of several tens to hundreds. We further investigate different halo gas pressures and energy injection methods.
Many of our simulated galaxies, but not all, develop bipolar outflows. We characterise the strength of the outflow by mass and energy outflow rates, and investigate the effect of changes to the details of the model. We find that supernovae are more effective if comprised into larger superbubbles. The weight \rev{and the pressure} of the halo gas is able to quench galactic outflows. 
\rev{The wind emerges from a series of superbubbles in regions where a critical star formation density is exceeded. The superbubbles expand into the gaseous halo at slightly supersonic speed, producing radiative shock waves with similar characteristics as the absorptions systems observed around LBGs.}

\end{abstract}


\section{Introduction}\label{intro}

Galactic winds are commonly characterised by supersonic, biconically shaped outflows perpendicular to the midplane of a galactic disc, energetic enough to carry significant amounts of gas far away from their host system, and in some cases even into the intergalactic medium (IGM, for a review see Veilleux, Cecil \& Bland-Hawthorn (2005)). Observations have revealed evidence for supersonic outflows in the spectra of nearby ultra-luminous infrared galaxies \citep[ULIRGs, Rupke, Veilleux \& Sanders 2002, 2005;][]{M2005} as well as in luminous $z>2$ infrared galaxies \citep{Sm2003,S2005}, Lyman break galaxies (LBGs) of $3<z<4$ \citep{L1997,P2000,P2001,P2002,A2003,Sh2003} and even in some gravitationally lensed Ly$\alpha$ emitters at $4<z<5$ \citep[ Frye, Broadhurst \& Ben{\'{\i}}tez 2002]{F1997}. The typical observational characteristics for galactic winds appear in most ULIRGs \citep[e.g.][]{H2000} and most of the LBGs at redshift $z \sim3-4$ \citep[e.g.][]{L1997}. It is obvious that they are common and may play an important role in galaxy evolution. Outflows on galactic scales can remove a substantial fraction of gas from a galaxy that would otherwise collapse into cold, dense molecular clouds and form stars \citep{SH2003,RT2006,S2006}. It has been pointed out that galactic winds can effectively carry metal-enriched gas from the galactic disc through the halo and straight into the IGM, hence being the main process responsible for the observed metallicity in the IGM \rev{\citep*[e.g. Tegmark, Silk \& Evrard 1993;][ but see also \citet{G1998,Silkea12}]{NT1997,H1998,DT2008}}.

\rev{The} exact mechanisms behind the driving forces of a typical wind are poorly understood. Observations \rev{of mainly nearby objects} show that the mass outflow rate is similar to the star formation rate (SFR), and that the efficiency of conversion of supernova (SN) energy into kinetic energy of the outflow is above 10 per cent \citep{VCBh2005}. The outflows are multiphase in nature and are limb brightened, suggesting a hollow cone structure \citep[e.g.][]{VCBh2005,SBh2010}.

SNe are generally believed to play a key role in providing the vast 
amounts of energy that can be observed in wind-like, but also convective 
galactic outflows. The SN energy is thought to 'thermalise', and generate a 
pressure-driven outflow similar to the stellar wind bubble \citep[e.g.][]{W1977}. 
A bubble produced by a single SN will expand due to its high internal overpressure, 
until the pressure has dropped to a value comparable to the bubble environment. 
Even before, the shock front produced by bubble expansion will collapse due to 
cooling, furthering the Rayleigh-Taylor instability, followed by a filamentary 
re-infall of cold, dense gas into the hot, over-pressured bubble region. 
By similar mixing processes, smaller bubbles are practically dissolved before 
they can reach the edge of the galactic disc. In order to drive a wind by pressure, 
it is therefore inevitable to create a sufficiently large region of thin and hot 
over-pressured gas, which harbours enough internal energy to provide a 
steady phase of expansion until the bubble can break out of the cold and 
dense gas disc \rev{\citep[e.g.][]{Mlea89,St2009,W2011}}. Simulations of 
the ISM \citep[e.g.][]{AB2004,AB2005} feature these overpressure regions 
within the hot phase of a multiphase ISM.  

\rev{Recent simulations have attempted to connect the ISM structure and dynamics 
to mass loss from galactic discs: Hopkins, Quataert \& Murray (2012) 
present galaxy scale simulations 
with parsec-scale resolution, including the effects of radiation pressure 
(local as well as long-range), stellar winds and supernovae. They show that different 
feedback processes are responsible for local and large-scale effects, leading
to an ISM structured on large scales into filaments and superbubbles. Spiral structuring is 
also important for the ISM in some of their models. They show that the mass-loss rates 
from the disc depend strongly on these details within the discs, which are therefore clearly 
a key issue in understanding the galactic wind phenomenon.  Similar to \citet{DVS2012}, 
who study mass loss from galaxy discs with SNe-feedback only, they do not 
model the interaction with the gaseous halo. Both, \citet{DVS2012} and \citet{Hopkea12} 
find mass 
outflow rates in excess of the star formation rate, in contrast to \citet{DT2008}, who do 
include the interaction with the gaseous halo (albeit with a simpler feedback model) 
and find mass loss rates of only a small fraction
of the star formation rate. This supports the conclusion of \citet{DT2008} that the interaction 
with the gaseous halo is another key factor in galactic wind studies. }

\rev{The energy input 
from a star-forming region extends over a few $10^7$ yrs \citep[e.g.][]{V2009}, 
which coincides with the estimated age of observed superbubbles \citep[e.g.][]{Bagetakea11}.
Superbubbles quickly come into pressure equilibrium with their surroundings, once 
the energy input ceases \citep[e.g.][]{K13}. This suggests that buoyancy has a potential to become important.
In elliptical galaxies, especially the central ones in clusters
  of galaxies, it is clear that light bubbles rise buoyantly through
  the hot atmosphere, dragging along cooler X-ray gas
  \citep[e.g.][]{Chu01,Formea07,Roedea07,Trembea12}, which is enriched
  in metals \citep[e.g.][]{Hea07,Kirkpea11}. For disc-dominated
  galaxies, which are the main focus of the present paper, the hot gas has
only been observed out to about one scale height, roughly 1-2~kpc,
above the disc \citep[compare the review by][]{Putmea12}. Hence, no
rising bubbles could be observed so far. From a hydrodynamics point of view
it is unclear if such bubbles should be present in disc-galaxy haloes:
the bubbles are likely jet-blown in the ellipticals, and would be related to
star-cluster outflows in the star-formation dominated disc
galaxies.
Also, stability is an issue \citep[e.g.][]{Kaisea05},
and it is currently unclear if the expected differences in Cosmic ray
content or magnetic-field dynamics \citep{GL12} would allow for
stably-rising bubbles in the haloes of disc galaxies. 
More likely, some mass entrainment and mixing takes
place \citep{Creasea13}, reducing the superbubble entropy, such that only lower halo
altitudes may be reached.
ISM structure, overpressure, turbulence and possibly buoyancy 
might therefore be considered to
be important factors for the onset and overall evolution of galactic
winds.}

\rev{Over recent years, a wealth of information has been collected specifically  for
  LBGs: Their masses are of order $10^9$-$10^{11} M_\odot$ \citep[e.g.][]{P2001,Steidea10}. 
The star formation rates are typically of order 10~$M_\odot$~yr$^{-1}$ and the star 
formation densities are of order 1~$M_\odot$~yr$^{-1}$~kpc$^{-2}$ \citep[e.g.][]{Erbea06}; 
the latter is similar to local starburst galaxies. 
LBGs are well-known for their associated absorption lines
\citep[e.g.][]{Kulea12,Lawea12}, commonly interpreted as expanding
(sometimes also infalling) shells of gas
\citep[e.g.][]{Rich03,K05,Tapkea07,Verhea08,Schaerea11}. 
Such shells may be interpreted as radiative shock waves related to a galactic outflow. 
For this to occur, the cooling time of the shocked gas must be below the one of the 
unshocked gas, which may be arranged for by a stratified gaseous halo. Because outflow velocities are found to be typically about 150-200~km~s$^{-1}$ \citep{Verhea08},
the sound speed in the gaseous haloes of LBGs may not exceed this value, limiting the halo temperature to $10^6$~K \citep[compare][]{K05}. It should also not be much lower, because otherwise one would also expect many slower shells. In this framework, the independence of the 
shell velocities from the star-formation properties \citep{Steidea10} 
may be understood in terms of a wind mechanism which is just able to power a weak shock, such that the outflow velocities of the shells are always constrained to be near the sound speed in the halo. 
The neutral hydrogen columns of the shells are between $2\times 10^{19}$~cm$^{-2}$ and 
$7\times 10^{20}$~cm$^{-2}$ \citep{Verhea08}. Because the shells consist of swept-up 
and cooled gas from the galaxy's halo, this number also indicates the column of hot gas 
before the passage of the shell. 
 }

\rev{Galactic winds} have recently been simulated by \citet{DT2008} and \citet{PSD2011}. \citet{DT2008} model their galaxies as cooling and collapsing Navarro-Frenk-White (NFW) spheres, and focus on the onset of a galactic wind working against the ram pressure of the in-falling halo material. They find that galactic winds arise only in low mass systems with comparatively small ram pressure which arises due to cooling and subsequent halo contraction, whereas larger ones will typically exhibit galactic fountains instead. \citet{PSD2011} study high-redshift galaxies ($z>9$) which are still in a phase of strong accretion by filamentary inflow of cold matter. They investigate if galactic winds may significantly alter the mass accretion rate of young galaxies in order to inhibit their further growth. They conclude that, though in these violently star-forming systems strong winds will develop, the accretion rate will not be affected, and hence there will be enough gas supply for long-lasting, intense star formation.
\rev{Most recently, \citet{Verhea12} have performed radiative transfer on hydrodynamics 
simulations similar to the ones in \citet{DT2008}. They find that line profiles with 
enhanced red wings, similar to outflowing shells, may also be produced from features of the 
ISM within or close to the galactic disk, as large-scale shells are not produced in the simulations. Interestingly, their result is strongly dependent on inclination, in stark contrast to observational results by \citet{Lawea12}. The latter might indicate that the underlying hydrodynamic model is not representative of the majority of LBGs.}

\rev{In these simulations, the simulated galaxies emerge self-consistently from some 
  physical input. While this is extremely useful for putting LBGs into the Cosmological context, there are apparently difficulties to connect to the observations. Here,  we use a fully tunable galaxy model as initial condition. We motivate the choice of our parameters from observations and Cosmological simulations and investigate the effect of the feedback strength and implementation. In particular, we make the important intermediate step to use superbubbles from star-cluster outflows instead of single supernova bubbles, and investigate the outflow properties as a function of the size of the superbubbles, thus parametrising the ISM physics
which leads to the development of different superbubbles in different types of galaxies. }

 In Section~\ref{sec:theo} we present theoretical considerations about the most likely 
wind drivers, involving an analytical model to sketch the processes during 
buoyancy-driven bubble expansion, the multiphase ISM and the energy requirements 
for kinetic wind driving. Section~\ref{sec:num} contains the simulation setup as 
well as the most 
important physics. We present our galactic wind simulations in Section~\ref{sec:results} 
and compare 
mass and energy outflow rates for different assumptions about the stellar feedback, 
and the thermal pressure of the gaseous halo. We find that galactic outflows are stronger 
for more concentrated supernovae, less halo pressure, and if we include a thermal 
energy component with the supernova events. We discuss these findings in Section~\ref{discu}.

\section{Theory}\label{sec:theo}

Here, we consider \rev{thermal energy injection, secular accumulation of kinetic 
energy and buoyancy} in detail, and in particular their relevance for driving galactic 
outflows. Since the energy balance of a galactic outflow depends critically on the 
surrounding halo, it is necessary to set up a self-consistent model of the latter in 
the first place.

\subsection{Setup}\label{sec:setup}
\subsubsection{Halo}\label{sec:halo}

LBGs with winds typically occur at redshifts between 3 and 4 (compare Section~\ref{intro}, above). Let us therefore consider a NFW halo in hydrostatic equilibrium at redshift $z=3.5$. The critical background density of baryons in the intergalactic medium (IGM), $\rho_{\rmn{crit,b}}$, can be obtained via 
\begin{equation}
  \rho_{\rmn{crit,b}}=\frac{3 \Omega_{\rmn{B}}H(z)^2}{8\pi G}
  \label{rhobg}
\end{equation}
(Ohta, Kayo \& Taruya 2003), where
\begin{equation}
  H(z)^2=H_0^2\left(\Omega_{\rmn{M}}(1+z)^3+\Omega_{\rmn{k}}(1+z)^2+\Omega_{\Lambda}\right),
  \label{hubble}
\end{equation}
with $\Omega_{\rmn{k}}=-0.02$. For a flat Lambda Universe it follows
from the Friedmann equations that
\begin{equation}
\Omega_{\rmn{M}}(z)=\frac{\Omega_{\rmn{M},0}}{\Omega_{\rmn{M},0}+\frac{1-\Omega_{\rmn{M},0}}{(1+z)^3}}.
\end{equation}
Using the present-day parameters of $\Omega_{\rmn{b},0}=0.044$ and $\Omega_{\rmn{M},0}=0.27$ one obtains $\Omega_{\rmn{M}}(z=3.5) = 0.97$ and $\Omega_{\rmn{b}}(z=3.5) = 0.16$, which, by combining equations~(\ref{rhobg}) and (\ref{hubble}), yields $\rho_{\rmn{crit,b}} = 1.4\times10^{-28}\,\rmn{g}\,\rmn{cm}^{-3}$.

By choice, the model system shall have a virial radius $r_{\rmn{vir}} = 25\,$kpc. With $r_{200}=0.94\,r_{\rmn{vir}}$ at $z=3.5$, this immediately yields a scale radius $r_{\rmn{s}} = r_{\rmn{vir}}/4 = 5.9\,$kpc by invoking a value for the concentration parameter $c_{200} = r_{200}/r_{\rmn{s}} = 4$, which is verified by \citet{Z2009} for our underlying redshift. The baryonic mass confined within $r_{200}$ may be pinned down via the critical baryon density, $\rho_{\rmn{crit,b}}$, to a value of $M_{200,\rmn{b}} = 2.2\times10^{10}\,M_{\odot}$.
As mentioned above, we assume that the gas in the halo is initially in hydrostatic equilibrium, and isothermal, suggesting a radially exponential distribution of baryonic matter:
\begin{equation}
  \rho_{\rmn{b}}(r,\theta) = \rho_{\rmn{crit,b}}\,\rmn{exp}\left(-\Phi_{\rmn{tot}}(r,\theta)\,\frac{0.59\,M_{\rmn{P}}}{k_{\rmn{B}}\,T}\right)
  \label{rhohalo}
\end{equation}
where $M_{\rmn{P}}$ is the proton mass, and
\begin{equation}
  \Phi_{\rmn{tot}}(r,\theta)=\Phi_{\rmn{disc}}(r,\theta)+\Phi_{\rmn{cent}}(r)+\Phi_{\rmn{NFW}}(r),
\end{equation}
with
\begin{equation}
  \Phi_{\rmn{NFW}}=-\frac{G\,M_{200}}{ r_{\rmn{s}}\,f(c_{200})}\,\frac{\rmn{ln}(1+r/r_{\rmn{s}})}{r/r_{\rmn{s}}}
\end{equation}
being the NFW potential dominating at larger radii. The other two potential components are due to the disc and the central bulge, respectively, and will be explained in sect. 2.1.2. The density distribution according to equation (\ref{rhohalo}) is visualised in Figure \ref{setup}. Note that the density in the inner parts of the halo remains within reasonable bounds due to our simulation domain being cut off at $r=0.4\,\rmn{kpc}$. Since the halo shall be isothermal, we can vary $T$ thus that the density $\rho_{\rmn{b}}$ at the inner edge $r=0.4\,\rmn{kpc}$ is not higher than typical disc density values, which are of order $10^{-24}\,\rmn{g}\,\rmn{cm}^{-3}$. The resulting temperature is $T=6.0\times10^5\,$K \rev{(close to the one inferred from observations compare Section~\ref{intro})}, and by integrating the now well-defined baryonic density profile, we obtain a baryonic mass of $1.0\times10^{9}\,M_{\odot}$ being situated in the hot halo. 
\rev{If this halo gas mass would be distributed across a wind shell of 10~kpc radius, it would produce a neutral hydrogen column of $10^{20}$~cm$^{-2}$, also what is required by the observations (compare Section~\ref{intro}). We have also checked that these parameters are in reasonable agreement with Cosmological simulations, which successfully reproduce observations of high-redshift galaxies \citep{SL2006,RS07,GS08,LS09,ST10}. From these simulations, we also find that the overall baryon fraction for comparable galaxies is always around or above the Cosmologically expected value, even for the strongest feedback. This is in agreement with a picture in which the so-called missing baryons are actually located in the gaseous haloes of galaxies \citep{SL2006,NSL2010}. For the Milky Way this picture has now been confirmed observationally \citep{G2012}. In the Cosmological simulations, we find about 10-20 per 
cent of the baryons in the hot halo. We have chosen here a smaller value of 5 per cent, 
still consistent with observations, which should favour the development of winds.
This would also account for some halo gas being lost already due to preceding
galactic wind activity.
\rvtwo{We also note that the stellar to halo mass ratio in our simulations, 7.3~per~cent,
is well within the range given by \citet{M13} for statistical halo abundance 
matching for our halo mass and redshift, 0.08 - 25~per~cent, but towards the high side 
of their central value, 1.4~per~cent.}
We discuss the implications of these choices in Section~\ref{discu}.}

The larger part of $M_{200,\rmn{b}}$, still amounting to $2.1\times10^{10}\,M_{\odot}$, must be considered to have settled into the disc. With the halo density $\rho_{\rmn{b}}$ given for all radii, the halo pressure $p$ can be obtained from the ideal gas equation
\begin{equation}
  p = n_{\rmn{b}} k_{\rmn{B}} T,
  \label{idealgas}
\end{equation}
where $n_{\rmn{b}}=\rho_{\rmn{b}}/(0.59 M_{\rmn{P}})$, due to ionisation. The initial equilibrium state for the halo will only hold as long as the temperature is kept constant. Yet since in some of our runs radiative cooling is permitted for the model halo, the subsequent temperature decrease leads to some contraction of the halo with time. This in some sense accommodates for the fact that galaxies at the given redshift are still accreting halo material in significant amounts. However, the interaction between (filamentary) infall of material into a galactic disc and the onsetting wind is beyond the scope of this work and is studied thoroughly by \citet{PSD2011}.

\begin{figure}
  \centering
  \includegraphics[width=.48\textwidth]{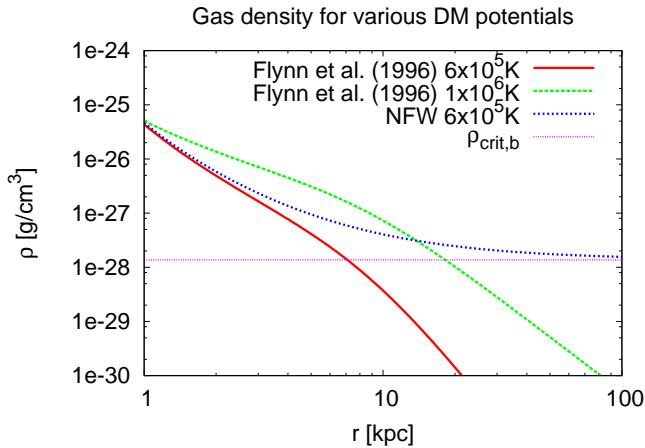}\\
  \vspace*{0pt}
  \caption{Hydrostatic gas mass density in $\rmn{g}\,\rmn{cm}^{-3}$ for three NFW haloes and two DM haloes with a $\rmn{ln}(r)$-profile as is described in \citet{FSC1996}, at different equilibrium temperatures, respectively. For the NFW profiles, density and therefore pressure converge against the cosmic background value quickly for every temperature, whereas this is not the case for the $\rmn{ln}(r)$-profiles. The polar angle for all curves is $\theta=\pi$.}
  \label{halodensities}
\end{figure}

The task of constructing an isothermal halo in hydrostatic equilibrium is encumbered by the condition that its density should converge against a certain background value.
\rem{ as described in section 2.1.1.} A halo potential of the form 
\begin{equation}
  \Phi_{\rmn{DM}}=\frac{v_{\rmn{rot}}^2}{2}\,\rmn{ln}\left(\left(\frac{r}{\rmn{kpc}}\right)^2+\left(\frac{r_{\rmn{s}}}{\rmn{kpc}}\right)^2\right)
\end{equation}
given by a constant rotational velocity $v_{\rmn{rot}}$ for large $r$ may seem physically justified. Such a model is described closer in Flynn, Sommer-Larsen \& Christensen (1996), however, this model entails the fact that the halo pressure will not converge. This means first of all that shock fronts could theoretically proceed to infinity as due to the resistant pressure decreasing strongly with $r$ they will accelerate forever. Furthermore, the density would have to drop adequately in order to maintain a constant temperature all over the halo, and would soon reach unreasonable values below the cosmic background (compare Figure~\ref{halodensities}). We hence adopted the NFW potential to overcome the described problems.

\subsubsection{Disc}\label{sec:disc}

\begin{table}
  \centering
  \begin{minipage}{75mm}
    \caption{External bulge and disc potential parameters for all simulations.}
    \begin{tabular}{@{}lrr@{}}
      \hline
      Component & Parameter & Value \\
      \hline
      Bulge & $r_{\rm{C_1}}$ & $1.35\,\rm{kpc}$ \\
      & $M_{\rm{C_1}}$ & $1.11\times10^{9}\,M_{\odot}$ \\
      & $r_{\rm{C_2}}$ & $0.21\,\rm{kpc}$ \\
      & $M_{\rm{C_2}}$ & $5.92\times10^{9}\,M_{\odot}$ \\
      \hline
      Disc & $b$ & $0.15\,\rm{kpc}$ \\
      & $a_1$ & $2.905\,\rm{kpc}$ \\
      & $M_{\rm{D_1}}$ & $2.442\times10^{10}\,M_{\odot}$ \\
      & $a_2$ & $8.715\,\rm{kpc}$ \\
      & $M_{\rm{D_2}}$ & $-1.073\times10^{10}\,M_{\odot}$ \\
      & $a_3$ & $17.43\,\rm{kpc}$ \\
      & $M_{\rm{D_3}}$ & $1.221\times10^{9}\,M_{\odot}$ \\
      & $r_{\rm{s,D}}$ & $2.05\,\rm{kpc}$ \\
      \hline
      \label{tab-discflynn}
    \end{tabular}
  \end{minipage}
\end{table}

Several approaches to establish a stable disc-halo system have been tested. A detailed description for a possible setup can be found in \citet{C2008}. In general, the following issues have to be kept in mind: Firstly, we want the gaseous disc to be rotationally supported (i.e. in hydrodynamic equilibrium), whereas the halo shall be pressure-supported (i.e. in hydrostatic equilibrium), which inevitably causes friction and shear effects in the transition zone. In addition, the halo cannot be truly set up in a pressure equilibrium with the disc, as the halo isobars are geometrically not parallel to those of the disc, which inevitably causes some motion in the halo. Therefore, we allow the system to relax for one Myr. The resulting setup is then sufficiently close to an equilibrium configuration to allow for the development of relatively stationary outflow solutions (compare below).
 As mentioned above in equation (\ref{rhohalo}), the total potential is built up of three components, where the disc component $\Phi_{\rmn{disc}}(r,\theta)$ is a combined form of a Miyamoto-Nagai potential \citep{MN1975}:
\begin{eqnarray}
  \Phi_{\rmn{disc}}&=&-\frac{G\,M_{\rmn{D_1}}}{\sqrt{R^2+\left(a_1+\sqrt{z^2+b^2}\right)^2}} \nonumber \\ && -\frac{G\,M_{\rmn{D_2}}}{\sqrt{R^2+\left(a_2+\sqrt{z^2+b^2}\right)^2}} \nonumber \\ && -\frac{G\,M_{\rmn{D_3}}}{\sqrt{R^2+\left(a_3+\sqrt{z^2+b^2}\right)^2}}.
\end{eqnarray}
The bulge component $\Phi_{\rmn{cent}}(r)$ is basically a central potential,
\begin{equation}
  \Phi_{\rmn{cent}}=-\frac{G\,M_{\rmn{C_1}}}{\sqrt{r^2+r_{\rmn{C_1}}^2}}-\frac{G\,M_{\rmn{C_2}}}{\sqrt{r^2+r_{\rmn{C_2}}^2}}.
\end{equation}
These two components are further described in \citet{FSC1996}, which we will use as the basic prescription for our disc setup. We have scaled down the mass-related parameters therein ($M_{\rmn{D_1}}$, $M_{\rmn{D_2}}$, $M_{\rmn{D_3}}$, $M_{\rmn{C_1}}$ and $M_{\rmn{C_2}}$) by a factor of 0.37 to match the residual disc mass (gas and stars) of $2.1\times10^{10}\,M_{\odot}$. The length-related sizes ($a_1$, $a_2$, $a_3$, $b$, $r_\rmn{C_1}$ and $r_\rmn{C_2}$) in the description by \citet{FSC1996} have been scaled down by a factor of 0.5 for our purpose, leaving our disc at a scale radius of $r_{\rm{s}}=2.05\,\rmn{kpc}$. An overview of all the values related to the bulge and disc potentials is given in Table \ref{tab-discflynn}. For comparison, a typical LBG is observed to have comparatively small size, and a mass probably an order of magnitude smaller (a few $10^{10} M_{\odot}$; see \citet{P2001}) than the more massive SINS galaxies \citep{G2008}. They form stars dominantly in a steady mode with a range of star formation rates, tens of solar masses per year not being uncommon \citep{P2001,Sh2003}.
\rev{With a gas fraction of 50 per cent in the disc, the star formation density 
ranges from 0.06 ~$M_\odot$~yr$^{-1}$~kpc$^{-2}$ to 1.4~$M_\odot$~yr$^{-1}$~kpc$^{-2}$ at the inner edge of the disc. The average value is 
0.3~$M_\odot$~yr$^{-1}$~kpc$^{-2}$, towards the lower end of the values reported 
by \citet{Erbea06}.}

\begin{figure}
  \centering
  \includegraphics[width=.48\textwidth]{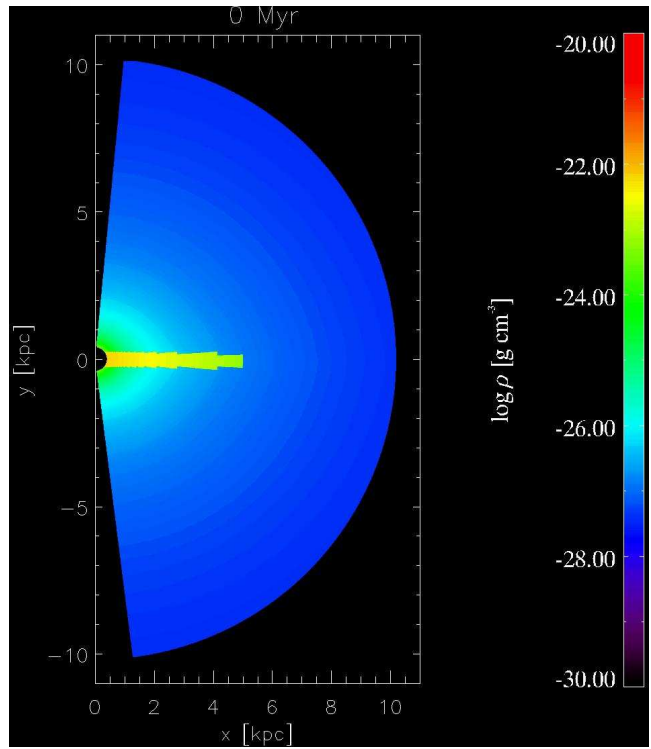}\\
  \vspace*{0pt}
  \caption{Initial mass density distribution in $\rmn{g}\,\rmn{cm}^{-3}$ at zero time. All simulations described herein are based on this setup. The disc is set up exponentially with $\rho_{\rmn{disc}}(r)\propto\rmn{exp}\,(-r)$, whereas the halo gas features an exponential-like distribution of $\rho_{\rmn{b}}(r,\theta)\propto\rmn{exp}\,(-\Phi(r,\theta))$. Note the slight deviation from spherical symmetry of the halo density due to the gravitational potential of the disc component.}
  \label{setup}
\end{figure}

\rev{Having} 50 per cent of the disc \rvtwo{mass locked in stars}
\rem{, which} gives us some freedom of choice for the gas density distribution, \rev{because} the disc potential is made up by the combined mass of gas and stars. We use an exponential (in radius) gas density profile with a cutoff at $r=5\,\rmn{kpc}$, that is vertically non-stratified. This latter fact is unproblematic since the disc will be given enough time to relax, so stratification will develop in the early course of the respective models ($\sim1\,\rmn{Myr}$). The disc density thus reads
\begin{equation}
\rho_{\rmn{disc}}(r,z)=\rho_{\rmn{disc}}(r)=\rho_{\rmn{disc},0}\,\rmn{exp}(-\frac{r}{r_{\rmn{s,D}}}),
\label{rhodisc}
\end{equation}
with $r_{\rmn{s,D}}$ being the disc scale radius, and $\rho_{\rmn{disc},0}=10^{-22}\,\rmn{g}\,\rmn{cm}^{-3}$. The disc has a vertical height of 500 pc, and thus the total gas mass is $1.1\times10^{10}\,M_{\odot}$, i.e. about 50 per cent of the mass implied by the disc potential. As an alternative to the exponential distribution, one could use a constant gas density profile, which has been observed e.g. by \citet{B2010} for NGC 2403; this is to be dealt with in a future paper. The disc gas pressure follows from the ideal gas equation~(\ref{idealgas}), just as for the halo gas pressure.\\
The gravitational force \rvtwo{is} accounted for by the implementation of $\Phi_{\rmn{tot}}(r,\theta)$ as an external potential. In Figure~\ref{setup}, the resulting density for our disc-halo system is shown as a contour plot; the initial and boundary conditions will be further explained in sect. 3.3. This setup condition applies to the complete set of simulations presented here and listed in Table \ref{tab-sims}.

\rev{\subsection{Overpressured superbubbles and the multi-phase 
interstellar medium}\label{sec:press}
Single SN bubbles dissolve into the surrounding ISM on a scale of order of
10~pc, which may also be derived from the LBG simulations presented below under 
the assumption of pressure equilibrium. This is a small value compared to the scale
height of disc galaxies (compare Section~\ref{intro}). Superbubbles from clusters 
of stars formed by the winds and SN explosions of many massive stars are therefore
more promising drivers of galactic winds. Superbubbles with energy requirements of 
tens to hundreds of SN and diameters of hundreds of parsecs are frequently found 
in nearby galaxies \citep[e.g.][also Section~\ref{intro}]{Bagetakea11}. 
Using such observed superbubbles directly in simulations also avoids the
possibly complex interaction of smaller interstellar bubbles leading 
to the emergence of the superbubbles in the first place \citep[e.g.][]{K13}.\\
We adopt this approach here. With our recipe for superbubble injection explained 
in Section~\ref{blastwave}, superbubbles with prescribed energetics are injected 
on a scale of about 100~pc. This leads to a dynamic, multiphase ISM, structured
on large scales with hot bubbles and cold filaments.
}

\subsection{Global kinetic energy}\label{sec:globkin}
\rev{A}   galactic wind could also be launched by the sheer amount of kinetic energy which accumulates within a gas-rich galactic disc over time. Let us \rem{therefore} consider a galactic gas disc with a mass of the order $10^{10}\,M_{\odot}$, as commonly found for Lyman-break systems. Let us further assume the SFR to be around $10\,M_{\odot}/\rmn{yr}$, which means in turn that we are going to encounter about one SN every 10 years. Normalised to the entire mass of the system this would mean a SN rate 10 times as high as in the Milky Way, which too has a gas mass of order $10^{10}\,M_{\odot}$ (the larger part of its mass is locked in stars) at a SFR of $1\,M_{\odot}/\rmn{yr}$. Our model system may therefore \rvtwo{be} regarded to be in a starburst phase. SNe are known to give rise to considerable turbulent motions within a disc (Dib, Bell \& Burkert 2006), each yielding a contribution of $\sim10^{51}\,$erg at a presumed efficiency $\epsilon=0.1$ to the overall kinetic energy stored within the gas phase of its host galaxy. Unlike internal energy, kinetic energy has the advantage that substantial fractions will not be radiated away immediately, but rather dissipate on the dynamical timescale \citep{Ml1998,B2006}. Allowing the turbulent energy to pile up for $\sim100\,\rmn{Myr}$ would result in an energy reservoir of order $10^{57}\,$erg for the disc as a whole. Since the gravitational binding energy is known to be of the same order for a $10^{10}\,M_{\odot}$ system of $5\,$kpc radial extent, material ejections from the disc into its surrounding galactic halo indeed becomes plausible at a certain point in time. The approach of launching a turbulence-driven outflow has been investigated \rem{closely} by \citet{SB2010}. In their models SN feedback was simulated by injecting unresolved kinetic energy, which is described by an isotropic pressure term in the Euler equation. Here, we also investigate kinetic energy driving, studying models where we inject only kinetic energy, instead of a combination of thermal and kinetic energy (compare below). We will however resolve the kinetic energy.


\subsection{Buoyancy of superbubbles}\label{sec:buoy}
\rev{Even large superbubbles may come into pressure equilibrium with their
 surroundings while still being close to or even within their galactic disc. In this case, 
buoyancy needs to be considered.}
An interesting physical quantity \rev{in this context}
is the entropy index $S$. Here, we calculate $S$ at relevant locations within the underlying NFW halo at redshift $z=3.5$. The entropy index is defined as
\begin{equation}
S=\frac{p}{n^{\gamma}},
\end{equation}
where $n$ is the number density of particles, the units of $S$ being given in keV cm$^2$. Generally, a bubble with an entropy index higher than its environment will experience a buoyant force, meaning that with S being known everywhere, we can easily determine the height a buoyant bubble can reach.\\
With the pressure expression from eq.~(\ref{idealgas}), the entropy index $S$ transforms into
\begin{equation}
S =\frac{k_{\rmn{B}}\,T}{n^{\gamma-1}}.
\label{entropy}
\end{equation}
Let us consider a bubble produced by a single SN in an early state of evolution. The entropy index is highest within the central hot gas phase of the bubble, and this is the region most relevant regarding buoyancy. Note that S is defined such that during the process of adiabatic expansion it is not going to change over time. For the hot bubble interior, S may decrease due to mixing and cooling. Cooling times are long compared to the simulation time, and mixing shall be neglected here in the first instance. This in turn means that the phase of evolution in which we investigate a bubble doesn't matter all too much. 
\rem{Since the rarefied, hot bubble interior has a very long cooling timescale, cooling is not significant here.}
A typical SN will release about $10^{51}\,\rmn{erg}$ of energy. From the equation of motion for a blast wave in the thin shell approximation, it follows that 60 per cent of this energy will be in the form of thermal energy. Implying an ejecta mass of 8$\,M_{\odot}$ and a bubble in an advanced state, e.g. with a radius of $10\,\rmn{pc}$ to start with, the density will be of order $0.1\,M_{\rmn{p}}\,\rmn{cm}^{-3}$. It follows then, assuming a temperature of $10^8\,\rmn{K}$, that the entropy index from eq.~(\ref{entropy}) reaches several $10\,\rmn{keV}\,\rmn{cm}^2$. Given a typical entropy index for the gas disc of order $10^{-4}\,\rmn{keV}\,\rmn{cm}^2$, the former value is certainly enough to raise the bubble away from the disc midplane into the disc-halo transition region. In our example, the values for $S(r_{\rmn{s}},\theta=\pi)$ and $S(r_{\rmn{vir}},\theta=\pi)$ in the halo amount to $9.1\,\rmn{keV}\,\rmn{cm}^2$ and $21.2\,\rmn{keV}\,\rmn{cm}^2$, respectively. Hence $S$ inside a bubble formed by several SNe will be typically high enough to exhibit buoyancy effects within the halo at least at low radii. This conclusion might however be affected by the (unknown) mixing of the different ISM phases. In our simulations, we include the buoyancy effect of the superbubbles. We inject the bubbles with even higher entropy index\rem{(compare Section~\ref{sec:bwinds} below)}, because numerical mixing - we have to inject the superbubble on a scale of a few grid cells - strongly reduces the entropy index. The energetic effect of buoyancy is however likely minor: 
\rem{While ascending, the bubble will vastly increase in size due to the radially exponentially decreasing environment pressure, thus allowing for its density to drop to negligible values compared to the inner halo environment. Because of the latter, buoyancy in the halo will likely affect only superbubbles in an advanced state of evolution, where their diameter has already grown up to the order of}
\rev{For a $100\,\rmn{pc}$ sized bubble and typical parameters of our simulations below, 
the acquired velocity would only be about 100~km~s$^{-1}$. Thus, the expected effect is 
that superbubbles first expand to pressure equilibrium and then hover near the disc-halo
interface. If fed sufficiently by other superbubbles, they may develop into a galactic wind. 
Otherwise, the bubble overshoots, collapses again and dissolves.}

\rem{. In that case, the acquired energy during ascension will be
\begin{equation}
E_{\rmn{buoy}} = \rho_{\rmn{halo}}\,V_{\rmn{bubble}}\,g_{\rmn{halo}}\,h,
\end{equation}
where $\rho_{\rmn{halo}}$ is the halo density, $g_{\rmn{halo}}$ its gravitational acceleration, $V_{\rmn{bubble}}$ the bubble volume and $h$ the height of ascension. In particular, a $100\,\rmn{pc}$-bubble will acquire some $10^{47}\,\rmn{erg}$ of energy while ascending $1\,\rmn{kpc}$. We shall keep these interim results in mind for comparison with our simulations.\\
\subsection{Multiphase interstellar medium}
As detailed in Section~\ref{sec:num} below, we allow radiative cooling only in between a certain range of temperatures. Below the lower temperature threshold for the cooling function, background radiation is assumed to keep the disc temperature stable at an overall value close to $10^4\,$K. The upper temperature threshold is a necessary tool in order to establish a resolved multiphase ISM. In reality, the ISM exhibits a filamentary structure, comprising cold ($T<200\,$K), dense filaments of molecular gas capable of producing stars, but only having a small volume filling factor of about 5 per cent \citep{AB2004}. The rest consists mainly of hot, rarefied gas with long cooling time, filling the large spaces in between the filaments. SN bubbles forming in this multiphase medium can expand to large radii, since the cooling process of their shock fronts, which would otherwise provide an efficient energy drain, cannot effectively take place within the hot, thin gas phase. In our numerical model we face a certain discrepancy in resolution: On the one hand, we are interested in a large-scale phenomenon (some kpc), inevitably meaning that the simulated domain, and therefore cell size has to be sufficiently large, on the other, a resolution of the multiphase gas disc structure would be desirable, calling for a cell size of order a parsec in order to resolve the cold, dense filaments, as is used by \citet{AB2005}. This is in particular important for shock fronts: The existence of a volume-filling low density component means that some part of the bubble expansion happens adiabatically. All these issues, however, enforce a compromise on our simulations to combine the key attributes of both phases; namely to provide the cells of a few dozen pc in length with the average gas density of the ISM, but still not allowing the cooling rate to be too efficient, which would be the case in rarefied, warm ISM regions represented by average-density cells. Note that with a standard cooling rate, a shock front within our disc can cool down to the environment value within even less than a time step, removing large quantities of energy from the SN bubble. A feasible solution is here to forbid cooling completely above a threshold temperature higher than the initial halo temperature but lower than the temperatures typically found in the SN bubble shock fronts. This procedure allows the bubbles to acquire a diameter well above the resolution limit before shell cooling sets in.}

\section{Numerical Methods}\label{sec:num}

We perform 3D simulations with the magnetohydrodynamics code NIRVANA \citep{ZY1997} on a spherical grid. 
\rev{Thus angular momentum is well conserved. In most runs, we simulate only a small fraction of the azimuthal angle, which allows to explore a larger part of the parameter space.}
\rem{\bf The choice of a spherical grid was made upon considering the simulation of a rotating system, as we found that conservation of angular momentum could be inaccurate and artifacts may originate on a Cartesian grid. The spherical grid has further been adapted for the reason that it allows us to simulate only an azimuthal wedge instead of a full disc, in order to keep computing time in reasonable bounds.} 
We have parallelised the code making use of the Message Passing Interface (MPI) library. Our simulations run for typically 48 hours on 6 SGI Altix processors. The gas evolution is calculated by solving the continuity, energy and Euler equations. A constant background gravitational potential accounts for the stellar and gaseous disc components, a bulge and the dark matter halo (compare above). The radiative cooling function used here is the equilibrium cooling curve described by \citet{SD1993}. It accounts for the overall metallicity which is assumed to be equal to the solar metallicity, and operates only within a temperature range between a lower limit of $10^4\,$K and an upper limit of about $10^6\,$K, with the exact value depending on the respective halo equilibrium temperature: For some of our models, the only effect of the upper cutoff is to prevent cooling in freshly injected SN shells. This is required to establish a more realistic SN remnant, before the shell cools and the remnant enters the snow-plough phase (see Sections 2.3 and 3.1 for more details). We also use the upper cutoff to entirely inhibit cooling of the halo in some simulations. We do this to account for the unknown halo metallicity, which has a strong impact on radiative cooling.  In this way, we cover the limiting cases of strong and negligible cooling of the halo.\\
We shall briefly describe the most important methods used in our studies and, if non-trivial, justify them physically; this includes cooling restrictions, SN triggering and their blastwave implementation.

\subsection{Stellar feedback}\label{sec:fb}

Star formation is triggered randomly for each single cell as soon as certain criteria are met, and SNe occur immediately in an amount related to the mass of stars produced. Star formation criteria include a local surface density exceeding the critical value \rev{$\Sigma_{\rmn{crit}}=10\,M_{\odot}$~pc$^{-2}$} required for star formation to set in \citep{K1998}. Before calculating the local surface density, a volume density criterion applies for each cell to ensure that it is part of a region dense enough to produce stars, which is, in particular, the disc. Cells having a density less than $2\times10^{-24}\rmn{g}/\rmn{cm}^3$ are considered to be either halo cells or too rarefied for star formation to set in. In a few special cases, large high-density gas regions can be found far away from the disc. We are to assume then that our model galaxy is essentially breaking up as a consequence of too strong feedback. In consequence, once the disc has lost integrity, the Kennicutt-Schmidt law might no longer apply. To avoid perturbations from this effect, the column from which the surface density is calculated, comprising only the aforementioned disc cells, shall be no higher than one fourth of the total $\theta$-range of the simulation domain.\rem{, or $0.21\,\pi$.} This value chosen here, however, is not a critical parameter. Finally, in order to allow the system some relaxation after setup, SNe shall not occur before $1\,\rmn{Myr}$.\\
In this model, we regard only SNe type II, since star forming galaxies are observationally dominated by this type. Given the Salpeter IMF for the stellar mass distribution, we can easily calculate that of $100\,M_{\odot}$ of gas locked up in stars, one type II SN progenitor exists, with the latter typically being as massive as $19.8\,M_{\odot}$ on average, considering stars within a range from 8 to 120 $M_{\odot}$.\\
We assume that stars in all our model galaxies form in accordance to a local Kennicutt-Schmidt law \citep{K1998}, given by
\begin{equation}
\Sigma_{\rmn{SFR}}=2.5\cdot10^{-4}\left(\frac{\Sigma_{\rmn{gas}}}{M_{\odot}\,\rmn{pc}^{-2}}\right)^{1.4}M_{\odot}\,\rmn{kpc^{-2}\,yr^{-1}},
  \label{KS}
\end{equation}
where $\Sigma$ denotes the respective surface densities for star formation and gas mass. The gas surface density $\Sigma_{\rmn{gas}}$ is calculated for every time step $\delta t$ and every grid point within the $r$-$\phi$-plane by integration of all disc cell masses along $\theta$ and dividing by the surface area $\delta r\cdot r\,\delta\phi$ of the respective column. Integrating along $\theta$ instead of the normal with respect to the disc midplane is a sufficient approximation since the disc extends only across a small angle $\delta\theta$. Moreover, constraining star formation by limiting the maximum column height to one fourth of the $\theta$-range of the simulation domain will ensure that the angle of integration is sufficiently small.\\
The SNe in our simulations are assumed to cluster in groups of 20 to 200 at a given time and place (henceforth referred to as a 'SN event'). Thus\rev{,} we can calculate a probability value for every disc cell and each time step, giving the likelihood for a SN event comprising $\zeta_{0}$ SNe,
\begin{equation}
  P_{\rmn{SN}} =\frac{\Sigma_{\rmn{SFR}}\,\delta r\cdot r\,\delta\phi\,\delta t}{100\,M_{\odot}\,\zeta_{0}\,n_{r,\phi}},
  \label{snprob}
\end{equation}
with $n_{r,\phi}$ being the number of disc cells in the respective range of integration along $\theta$. $\zeta_{0}$, referred to as the 'event size' herein, is a preset parameter which will be kept constant during each single simulation. A random number is then drawn for each disc cell at every time step. The occurrence of a SN event is then triggered according to the local probability. Note that any altering of the resolution has to come with an appropriate change in the event size range; too high event sizes will require a manifold of the gas mass available in the cell, too small event sizes may produce unresolved bubbles.

\subsection{Blast wave implementation}\label{blastwave}

In this subsection we provide details of our blast wave implementation and follow the evolution of a single superbubble in a test simulation. If a SN event is determined to occur for a specific cell, the following modifications in mass and energy will immediately take place: An amount of $100\;\zeta_{0}\;M_{\odot}$ is regarded to be no longer available in gaseous form since it is bound in stars, and henceforth removed from the cell. This amount can, in case of \rvtwo{large $\zeta_{0}$, exceed} the cell mass, however, within our range of event sizes the total mass deficit due to this error is below a five per cent threshold for the entire disc, and hence considered negligible as it will insignificantly alter the disc dynamics. 25 per cent of the newly formed stellar mass is returned to the gas phase due to stellar winds and SN ejecta. So, essentially, our code removes $75\,\zeta_0\,M_{\odot}$ of gas from the SN-triggering cell. The remaining mass is distributed equally among the six neighbouring cells except for a small remainder of $10^{-28}\,$g/cm$^3$ within the central cell, so that the density increase is the same in all six adjacent cells. We assume an energy injection of $10^{51}\,\rmn{erg}$ per SN. 
\rev{According to the thin shell approximation for blastwaves with instantaneous 
energy injection,}
60 per cent of this energy is released as internal energy, fed into the SN cell and thus building up an overpressure with respect to the surroundings. The remaining 40 per cent of the energy total is kinetic energy, added as an extra velocity component to the neighbour cells. This velocity of the SN ejecta, $v_{\rmn{ex}}$, is typically greater than $10\,\rmn{km}\,\rmn{s}^{-1}$ upon release, and therefore supersonic with respect to the sound speed inside the dense disc material, in agreement with superbubble observations \rev{\citep[e.g.][]{Bagetakea11}}.
\rev{The temperature in the SN cells varies due to the spherical geometry of the grid. It is however always of order $10^{10}$~K or higher, exceeding the value determined by 
\citet{DVS2012} to achieve a converged outflow behaviour, $10^{7.5}\,\rmn{K}$.}
\rev{While this high temperature prevents cooling in the bubble interior, cooling is still very efficient in the shocked shell of the injected superbubble: In an interstellar bubble, the energy lost by adiabatic expansion from the bubble interior is used to accelerate the surrounding shell, which drives a shock into the ambient medium, where the energy finally thermalises and radiates. The total energy of a superbubble is lost within about $10^6$~yrs, once the energy input has stopped \citep{K13}. Thus, in order to establish superbubbles of hundreds of parsecs diameter on the grid, as observed, it is necessary to suppress cooling in the shocked superbubble shell 
for a short time interval after injection of the superbubble. This is not unphysical, since 
our simulations are not meant to explain the origin of the ISM structure, for which radiation pressure and stellar winds are crucial \citep{Hopkea12}, but to explore the effects of superbubbles with given sizes. We implement this by a threshold value above which no cooling is taking place. For most simulations, we have chosen this threshold value to be $10^6$~K, slightly above the halo temperature. In this case, no cells apart from the shells of freshly injected superbubbles are affected. Only in simulations with non-cooling gas haloes we artificially suppress cooling of the halo gas by setting the threshold to a value slightly below the halo temperature, in order to also suppress cooling in the halo gas. This initialises slightly bigger superbubbles in the latter runs. But the effect is shown below not to be significant, as the mass and energy loss rates are higher for the case when the halo is allowed to cool.  }

\rvtwo{The choice of $10^6$~K for the threshold value implies a critical 
shell-expansion velocity of 271~km/s, above which the expanding shells of freshly injected bubbles are assumed here not to cool. Hence, cooling is allowed way before 
a superbubble reaches observed velocities \citep[$\lesssim 30$~km~s$^{-1}$,][]{Bagetakea11}. A much higher threshold value (say factor of ten) would push the critical shell velocity up, well into the regime of single-supernova shells, which are not addressed in our simulations. If applied to our superbubble setup anyway, the injected energy would be radiated away quickly, and the bubbles would dissolve without producing any large-scale effects. Because for blastwaves the postshock temperature depends on the radius to the third power, smaller ($\approx$~factor of two) changes in the threshold temperature do not change the physics significantly. We have determined the threshold value experimentally to ensure that the superbubbles are just established properly on the grid. The exact choice of the threshold value therefore has an effect on the net energy input. However, we do not model 
the emergence and early evolution of the superbubbles 
\citep[compare e.g.][for a discussion]{K13}, but use idealised superbubbles instead.
We study the effect of variations of the superbubble-injection mechanism in 
Section~\ref{sec:kwinds}. However,
instead of varying the threshold temperature, we directly vary the amount of 
injected energy per bubble.}

\begin{figure}
  \centering
  \includegraphics[width=.48\textwidth]{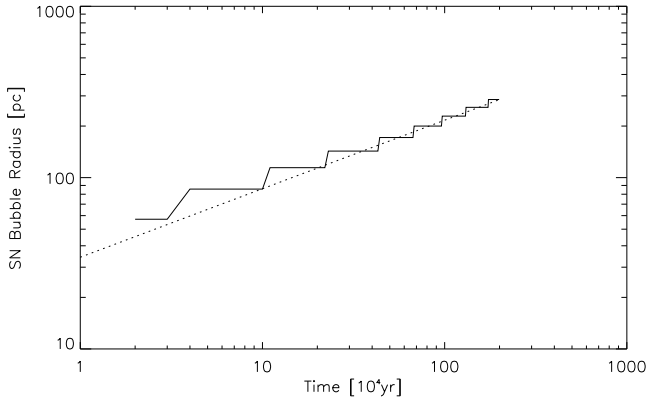}\\
  \vspace*{0pt}
  \caption{SN blast wave expansion of a $10^{53}\,\rmn{erg}$ event in an isotropic $10^{-24}\,\rmn{g}\,\rmn{cm^{-3}}$ medium. As expected, the blast wave expands in good agreement to a $r\sim t^{0.4}$ law (dotted line).}
  \label{shockbox}
\end{figure}

\begin{figure}
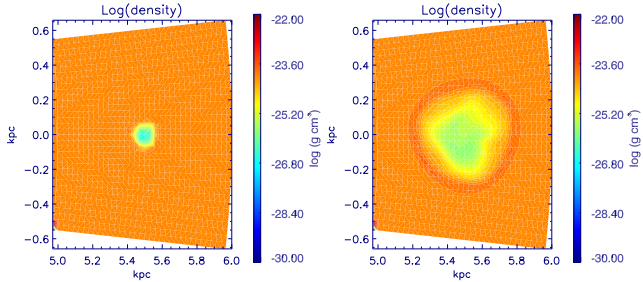

  \centering
  \includegraphics[width=.236\textwidth]{lgd010.epsi}
  \includegraphics[width=.236\textwidth]{lgd199.epsi}
  \vspace*{0pt}
  \caption{Mass density 0.1 Myr (left) and 2.0 Myr (right) after energy release.}
  \label{boxdensity}
\end{figure}

To check the behaviour of our blast wave implementation, we have modelled a box of $35^3$ cells on a spherical grid section of $5\,\rmn{kpc}<r<6\,\rmn{kpc}$, and $0.07\pi$ in each $\theta$ and $\phi$ direction. There is neither an external potential, nor does any other force (e.g. centrifugal) apply. The overall density is set to $\rho=10^{-24}\rmn{g}\,\rmn{cm}^{-3}$, and the temperature to $10^4\,\rmn{K}$ which is a common value for disk material in the models presented below. An energy equivalent of 100 SNe, or $10^{53}\,$erg, is released at $t=0$ right in the centre of the box as described above, forming an over-pressured, expanding hot gas bubble within a few 10,000 years (Figure~\ref{boxdensity}). There is no cooling taking place in this test run. We follow the bubble expansion over $2\,\rmn{Myr}$, tracing the distance between the shock front and the centre of explosion (Figure~\ref{shockbox}) as well as the energy decrease with time. The shock front in the underlying model is found to expand with a $r(t)\sim t^{0.4}$ law, as expected. Note however, that this is the expansion behaviour as expected from a bubble produced by one single SN. Superbubbles powered by many SNe spread out in time should rather expand with $r(t)\sim t^{0.6}$ \citep{O2009}. This is because all of our bubble-producing SNe are triggered in one cell within one time step, as resolution prevents us from spreading SNe reasonably in space and time, in order to produce more realistic superbubbles. We estimate that this effect increases our bubble sizes artificially by about 25 per cent despite the smaller expansion rate, as we start out with a much higher energy. On the other hand, we find that about 10 per cent of the initially released energy is lost by numerical effects within the first 100,000 years, however, any further loss thereafter is comparatively small. Because the advection step of our code conserves only the thermal and not the kinetic energy exactly, preferably the kinetic energy component will be lost. Figure~\ref{boxdensity} shows two snapshots of the SN bubble evolution, respectively 0.1 and 2 Myr after the event was triggered; the inner, rarefied region carries the internal energy. The kinetic energy resides within the compressed high-density region surrounding the bubble. Its slightly asymmetric form and imbalances in the kinetic/thermal energy distribution are a result of the coarse implementation.

\begin{figure}
  \centering
  \includegraphics[width=.48\textwidth]{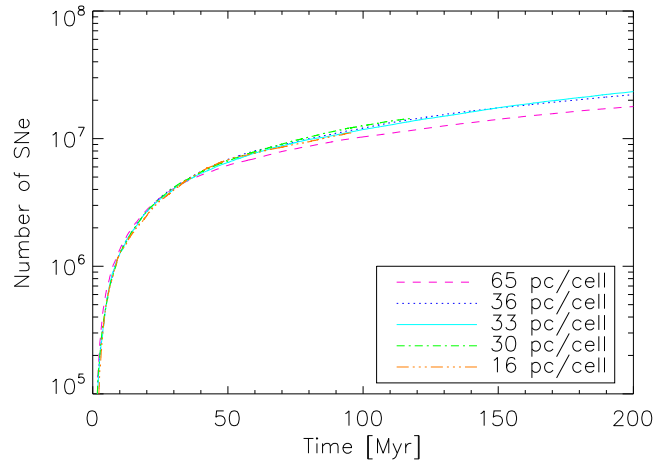}\\
  \vspace*{0pt}
  \caption{Cumulative supernova rate at a given time for our standard run at different resolutions (R16 - R65, compare Table \ref{tab-sims}). \rvtwo{The resolution given in the inset legend refers to the uniform radial resolution of a given run.}}
  \label{SFRres}
\end{figure}

\subsection{Setup, boundary and initial conditions}\label{sec:ini}

\begin{table*}
  \centering
  \begin{minipage}{100mm}
    \caption{Simulation parameters.}
    \begin{tabular}{@{}lcrrrrccc@{}}
      \hline
       & \multicolumn{3}{c}{Resolution\footnote{For the angular coordinates, the resolution at the inner and outer radial boundary is given.}} & \multicolumn{2}{c}{SN energy\footnote{The energy released by a SN event is subdivided into a kinetic and a thermal component.}}
      & Event size & Res.\footnote{The term 'Residual density' refers to the density left over in a cell after being subject to a SN event.} & Halo\\
      Run& $\delta r$ & $ r\,\delta \theta$ & $ r\,\delta \phi$ 
      & $E_{\rmn{kin}}$ & $E_{\rmn{therm}}$ & $\zeta_0$  & density& cooling \\
      & pc&pc&pc & \% & \% & SNe & g~cm$^{-3}$ & \\
      \hline
      R16 & 16 & 6-149 &6-124 &$40$ & $60$ & $100$  & $10^{-28}$ & yes \\
      R30 & 30  & 11-279 &9-233&$40$ & $60$ & $100$  & $10^{-28}$ & yes \\
      R33 & 33 & 12-307& 10-256 & $40$ & $60$ & $100$  & $10^{-28}$ & yes \\
      R36 & 36  & 13-335 & 11-279 & $40$ & $60$ & $100$  & $10^{-28}$ & yes \\
      R65 & 65  & 24-605 & 20-504& $40$ & $60$ & $100$  & $10^{-28}$ & yes \\
\rev{ STaz} & 33  & 12-307& 10-256& $40$ & $60$ & $100$  & $10^{-28}$ & yes \\
\rev{ STpol} & 33  & 12-307& 10-256& $40$ & $60$ & $100$  & $10^{-28}$ & yes \\
      ST20 & 33  & 12-307& 10-256& $40$ & $60$ & $20$  & $10^{-28}$ & yes \\
      ST50 & 33  & 12-307& 10-256& $40$ & $60$ & $50$  & $10^{-28}$ & yes \\
      ST100 & 33  & 12-307& 10-256& $40$ & $60$ & $100$  & $10^{-28}$ & yes \\
      ST200 & 33  & 12-307& 10-256& $40$ & $60$ & $200$  & $10^{-28}$ & yes \\
      KE0.4 & 33  & 12-307& 10-256& $40$ & $0$ & $100$  & $10^{-28}$ & yes \\
      TE0.6 & 33  & 12-307& 10-256& $0$ & $60$ & $100$  & $10^{-28}$ & yes \\
      TE0.4 & 33  & 12-307& 10-256& $0$ & $40$ & $100$  & $10^{-28}$ & yes \\
      B100 & 33  & 12-307& 10-256& $40$ & $60$ & $100$  & $10^{-27}$ & yes \\
      NC20 & 33  & 12-307& 10-256& $40$ & $60$ & $20$  & $10^{-28}$ & no \\
      NC50 & 33  & 12-307& 10-256& $40$ & $60$ & $50$  & $10^{-28}$ & no \\
      NC100 & 33  & 12-307& 10-256& $40$ & $60$ & $100$  & $10^{-28}$ & no \\
      NC200 & 33  & 12-307& 10-256& $40$ & $60$ & $200$  & $10^{-28}$ & no \\
      \hline
      \label{tab-sims}
    \end{tabular}
  \end{minipage}
\end{table*}

We run our simulations on a 3D spherical grid, with the radial dimension $r$ extending from 0.4 to 10.2 kpc, the polar angle $\theta$ covering a section between $0.04 \pi$ and $0.96 \pi$, and the azimuthal angle $\phi$ covering only a narrow 'wedge' of the disc within $-0.04 \pi$ and $0.04 \pi$ in range. Note that the space close to $\theta=0$ and $\theta=\pi$ as well as the one at $r<0.4\rmn{pc}$ must be omitted, as due to the spherical geometry grid cells within this space would become increasingly narrow. This in turn would lower their crossing timescales significantly, requiring high computing times for the innermost zones. The simulation of just a small azimuthal sector of the disc instead of the whole $\phi$-range 
\rev{implies large-scale rotational symmetry. We have also performed control runs 
relaxing the  assumptions about the azimuthal extent (STaz) and the polar-axis 
cutout (STpol). The polar axis cutout affects the energy fluxes by about 10~per~cent.
Otherwise, the effects are minor and are discussed in the appendix.}

For the main runs, the simulation domain is divided into $300\times96\times10$ grid cells in $r-$, $\theta-$ and $\phi-$directions, respectively. Thus, a region near the disc midplane at $r=1\,\rmn{kpc}$ is spatially resolved to $\sim33\,\rmn{pc}$ 
\rvtwo{(compare Table~\ref{tab-sims} for details)}. Our choice made here concerning the resolution will be explained in more detail at the end of this section. We choose reflective boundary conditions for the lower $r$ boundary and the upper and lower $\theta$ boundaries each, whereas on the outer boundary in $r$-direction inflow and outflow of material shall be permitted. The boundary conditions for the boundaries in azimuthal direction ($\phi$) are chosen to be periodical. Table \ref{tab-sims} shows the total set of simulations performed in the frame of this work with their respective parameters.\\

\subsection{Resolution}\label{sec:res}

We have investigated the resolution dependence of the SFR (see Section 5.1 below for a discussion of the dependence of the outflow rates on resolution), varying the reference resolution at $r=1\,\rmn{kpc}$ radius from 16 pc to 65 pc (R16 - R65, compare Table \ref{tab-sims}) for a standard simulation.
Assuming one SN in 100 $M_{\odot}$ of stars formed, we find the SFR in our $10^{10}\,M_{\odot}$ system at all resolutions to be about $10\,M_{\odot}\,\rmn{yr}^{-1}$, yielding an SFR per unit mass of $10^{-9}\,\rmn{yr}^{-1}$. As a comparison, this is several ten times the SFR per unit mass in the Milky Way, which would be of a few $10^{-11}\,\rmn{yr}^{-1}$. Our SFR is therefore in the relevant range; e.g. \citet{P2001} observe values of about $10-70\,M_{\odot}\,\rmn{yr}^{-1}$ for their sample of $10^{10}\,M_{\odot}$-LBGs at redshift $z\sim3$, which, accordingly, would result in an SFR several $10^{-9}\,\rmn{yr}^{-1}$ per unit mass (or a few $10^{-1}$ SNe per year). The overall SN rates of our model galaxy are displayed in Figure~\ref{SFRres} for all resolutions. The graph for $65\,\rmn{pc}$ resolution shows the strongest deviation, indicating that too coarse resolutions will notably affect the star formation rate. All graphs agree within 26 per cent, however, if we regard only resolutions of $36\,\rmn{pc}$ and finer, the error reduces to nine per cent.

\begin{figure*}
  \centering
  \includegraphics[width=0.93\textwidth]{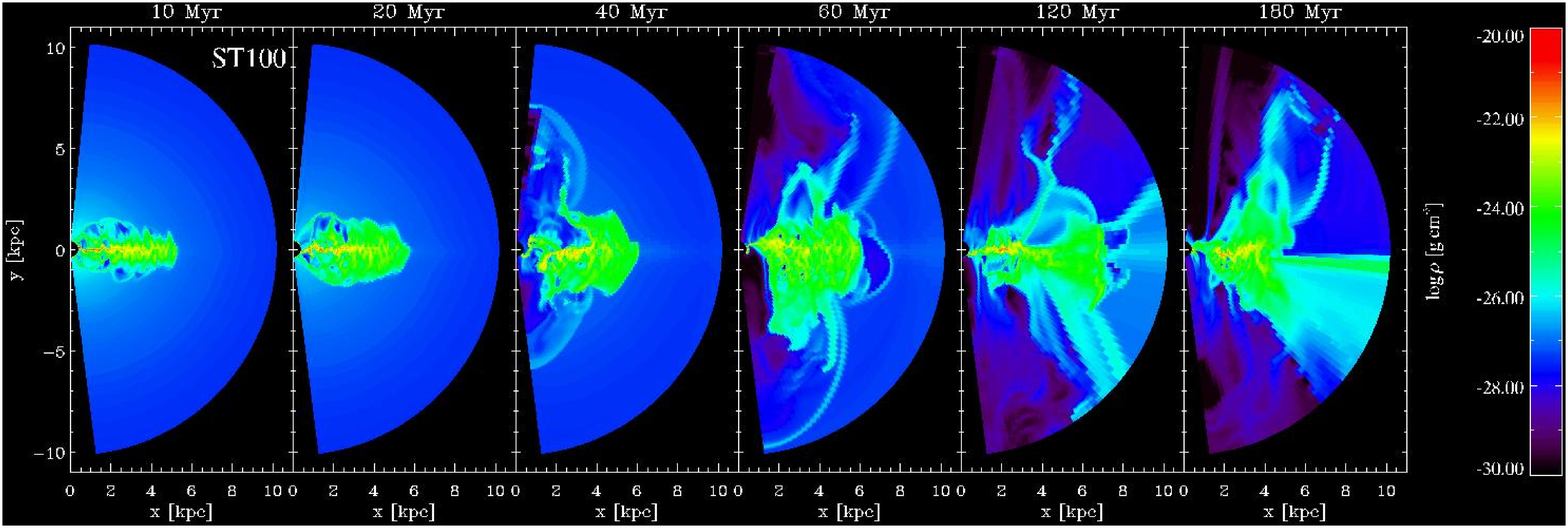}\\
  \includegraphics[width=0.93\textwidth]{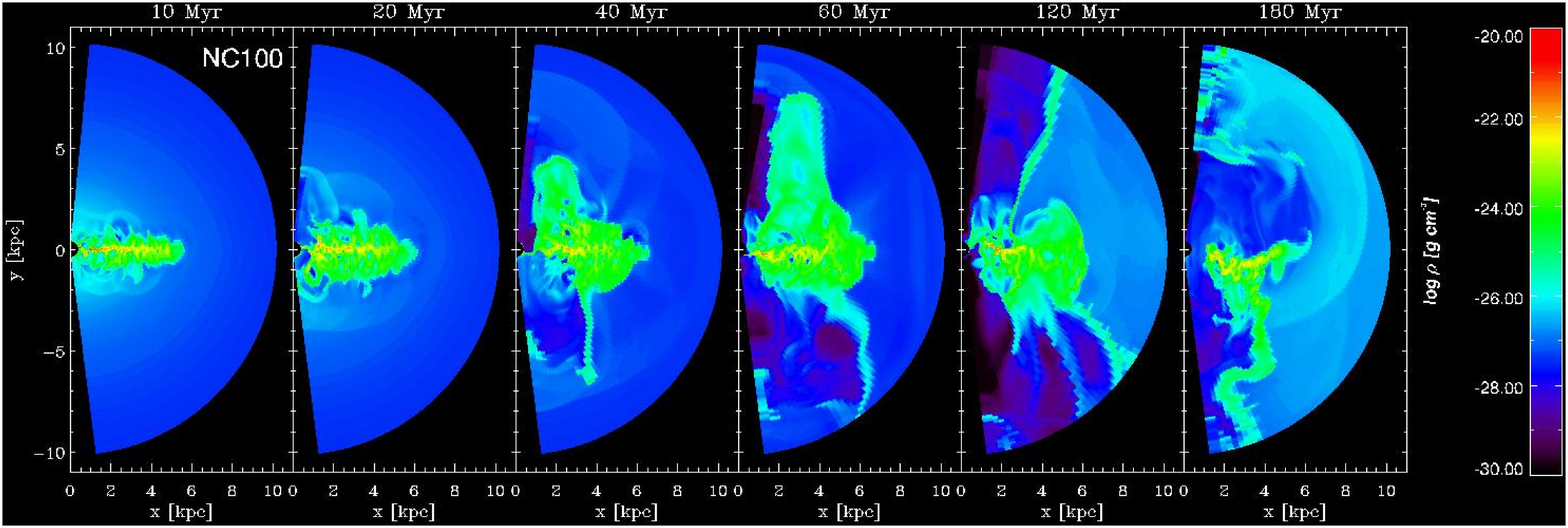}
  \vspace*{0pt}
  \caption{$Top$: Simulation ST100 with each SN releasing $4.0\times10^{50}\,\rmn{erg}$ as kinetic, and $6.0\times10^{50}\,\rmn{erg}$ as thermal energy. Note that the time span between two snapshots is not always the same; the elapsed time is denoted above each snapshot. Shown is the logarithm of the density in meridional midplanes. $Bottom$: Simulation NC100 with a non-cooling halo; see subsection 4.4. for details.}
  \label{lgd_ST100}
\end{figure*}

\section{Results}\label{sec:results}

We begin with an investigation of how the method of SN energy injection affects the emerging wind. For this purpose, we have run a set of simulations with $\zeta_0=100$. One simulation uses the Sedov-Taylor blast wave model, and hence both kinetic and thermal energy are injected with every SN event (denoted 'ST100'). In addition, two models were calculated, injecting a purely thermal energy fraction of 40 per cent (denoted 'TE0.4'), and 60 per cent ('TE0.6') of the total SN energy yield, respectively, and another one, injecting a purely kinetic energy fraction of 40 per cent ('KE0.4'). The characteristics of the pressure-driven and the kinetic energy driven cases are discussed in the first two subsections, respectively.\\
\rem{The third subsection includes an analysis of the contribution of buoyancy to the wind energy in ST100, which will be compared to our theoretical consideration in section 2.2.\\}
All the runs presented in subsections~\ref{sec:pwinds} and \ref{sec:kwinds} \rem{and \ref{sec:bwinds}} 
include a cooling halo. Since halo pressure is reduced by cooling, winds will arise comparatively easily in this case, allowing for more prominent effects more suitable for later comparison. Subsection~\ref{sec:bubsize} investigates the question how the sizes of SN bubbles can affect the strength of galactic winds; for this we have run another set of three simulations featuring Sedov-Taylor blast wave models and different event sizes each. In contrast to the previous runs, the runs in subsection~\ref{sec:bubsize} are each performed twice, with both, a cooling and a non-cooling halo, respectively, to investigate the limiting cases of the possible effects of varying metallicities in such objects. We show that the different halo pressures have a significant effect on the wind. All of our results herein will then be compared in the final subsection.

\begin{figure}
  \centering
  \includegraphics[width=.46\textwidth]{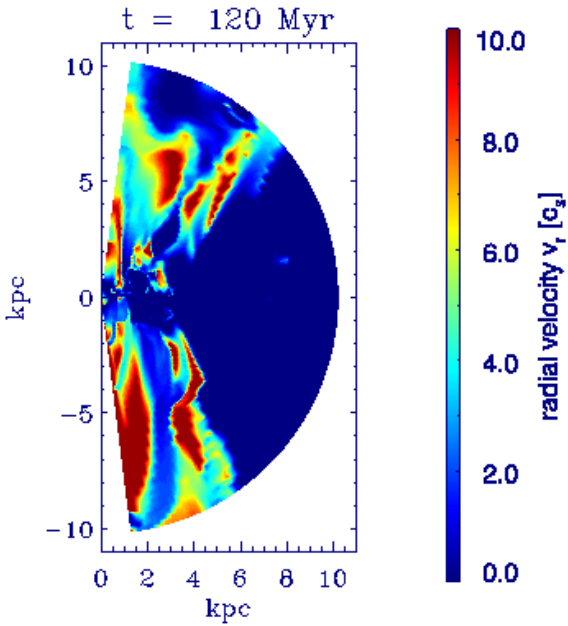}
  \vspace*{0pt}
  \caption{Radial velocity of outflowing gas regions in units of their respective local sound speed $c_{\rmn{s}}$ for simulation ST100. The velocities are capped at $0\,c_{\rmn{s}}$ and $10\,c_{\rmn{s}}$.}
  \label{vel_ST120}
\end{figure}

\subsection{Pressure-driven winds}\label{sec:pwinds}

In Figure~\ref{lgd_ST100} we show the mass density distribution of simulation ST100 at times of 10, 20, 40, 60, 120 and $180\,\rmn{Myr}$. We can clearly discern individual superbubbles expanding already at $10\,\rmn{Myr}$ beyond a height of $1\,\rmn{kpc}$ above and below the disc. 
\rev{At this point they have expanded out to pressure equilibrium and hover above and below the disc due to 
the long buoyant rise time. Bubbles in the outer part of the disc collapse back to the disc.
Bubbles inside a radius of 3~kpc however merge and are fed sufficiently 
quickly to prevent them falling back to the disc. The region of the disc where 
this happens has a star formation density greater than about 0.1~$M_\odot$~yr$^{-1}$~kpc$^{-2}$.}
These bubbles keep expanding, driven by their overpressure against the radially quickly declining halo pressure. At $40\,\rmn{Myr}$ the superbubbles 
\rev{create} a low density funnel close to the axis of symmetry. Since the gas inside this structure provides less resistance to subsequently escaping superbubbles than the rest of the halo region, material from the succeeding bubbles will continue to flow at ease through the funnel. 
\rev{The individual expanding superbubbles are however still identifiable by individual shock fronts, which may be more easily seen in the accompanying movie of run STaz. The funnel}
is surrounded by a conical structure of notably denser material which was originally entrained from the dense disc by outgoing shock fronts and hence continues to move outwards. Over time, enormous amounts of SN energy are fed into the disc, which in turn becomes extremely turbulent: large portions of gas are torn out of the disc midplane, partially due to entrainment by the wind, but eventually fall back. The shape of the disc gets highly irregular and clumpy but the disc remains overall intact.\\
Since we are dealing with a rather massive system, it might seem likely, regarding the studies by \citet{DT2008}, that outflows appear preferably in the form of galactic fountains. We make here the usual distinction \citep[compare e.g. ][]{DT2008} between the two common types of outflow solutions: A \textit{wind} is defined to be supersonic with respect to its internal sound speed. A \textit{fountain}, on the other hand, is subsonic. Galactic fountains are therefore much more susceptible to the Kelvin-Helmholtz instability and usually turbulent. Both types of solutions may in principal be bound to the galaxy or reach escape velocity. The smaller bulk velocity of the fountain solution usually prevents it from escaping the galaxy and the flow becomes convective, lead by a roughly spherical weak shock or sound wave around the whole system.\\
In contrast, the bulk velocities in the wind gas may easily reach escape velocity. Due to the geometrical constraint from the galactic gas disc, the outflow becomes conical. Figures \ref{lgd_ST100} and \ref{vel_ST120} demonstrate that the outflow which has emerged in run ST100 has developed all the usual characteristics for a wind solution. The escape velocity at 10 kpc distance from the disc amounts to $v_{\rmn{esc}}=426\,\rmn{km}\,\rmn{s}^{-1}$, which is well below the typical wind velocities close to $10^3\,\rmn{km}\,\rmn{s}^{-1}$.
The difference to \citet{DT2008} is mainly the size of the disc. \citet{DT2008} have chosen a much larger disc and therefore might not reach the required SN density to drive the outflow.

\subsubsection{Mass outflow}\label{sec:mdot}

\begin{figure}
  \centering
  \includegraphics[width=.48\textwidth]{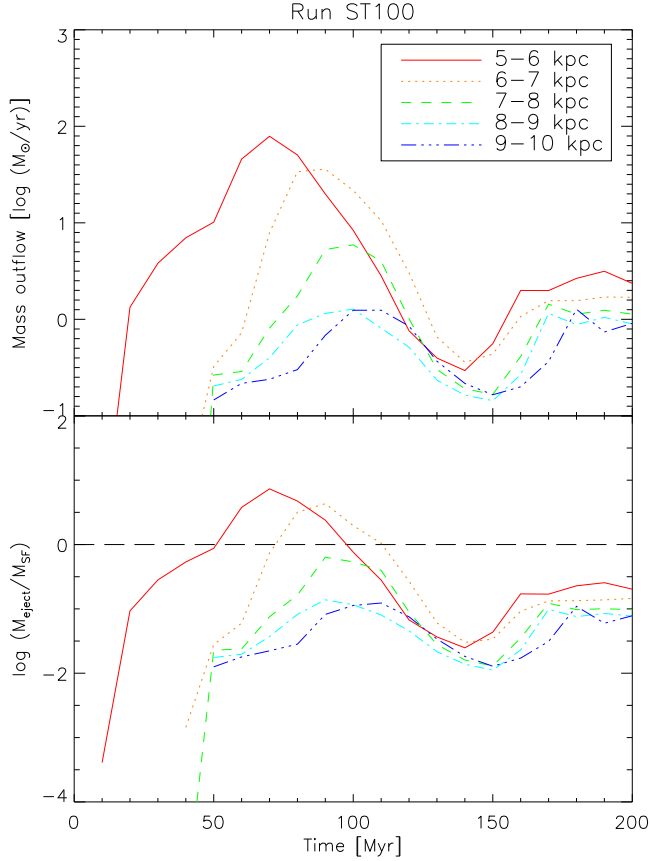}
  \vspace*{0pt}
  \caption{Analysis of simulation ST100. $Top:$ Mass flux rates through different shells of respective thickness $1\,\rmn{kpc}$. $Bottom:$ Efficiency of mass output, defined as the ratio of outflowing mass $M_{\rmn{eject}}$ to star-producing mass $M_{\rmn{SF}}$. The dashed black line marks unity.}
  \label{massflux_ST100}
\end{figure}

For a quantitative analysis of our models, we calculate the net mass flux across a spherical shell of inner radius $r_{\rmn{i}}$ and outer radius $r_{\rmn{o}}$. We start with 
\begin{eqnarray}
l_{M}(r,t)&=&k_{\phi}\frac{1}{\Delta r}\int_{0.04\pi}^{0.92\pi}\int_{-0.04\pi}^{0.04\pi}\int_{r_{\rmn{i}}}^{r_{\rmn{o}}}\rho(r,\theta,\phi,t)\,v_r(r,\theta,\phi,t) \nonumber \\ && \rmn{d}\theta\,\rmn{d}\phi\,r^2\,\rmn{sin}\theta\,\rmn{d}r,
  \label{massflux_eq}
\end{eqnarray}

which is the net mass flux at any point $t$ in time for a spherical layer of grid cells at a given radius $r$. The factor $k_{\phi}=25$ is a correction term which accounts for the fact that our simulation box covers only $1/25$ of the total $\phi$ range. Due to the box limits in $\theta$ range, a part of the wind at the poles is neglected. Due to the small surface area, this error is not significant \rev{(of order 1~per cent, compare appendix)}. The average mass flux for all layers at radii $r_{\rmn{i}}<r< r_{\rmn{o}}$ is determined every 1 Myr, and then averaged over 10 Myr, yielding the total net mass flux $L_{M}=\left<l_{M}(r,t)\right>$. Figure~\ref{massflux_ST100} shows mass flux rates from 0-200 Myr for run ST100 across shells of respective thickness of $\Delta r=1\,\rmn{kpc}$ for various shell positions. In the innermost shells, winds show up earlier and stronger, however, a large fraction of the outflowing mass in these inner shells is likely to represent entrained disc material. This material might, in some cases, fall back soon after its ejection from the disc, and actually not contribute to the mass carried away by the wind.

\subsubsection{Energy outflow}\label{sec:edot}

\begin{figure}
  \centering
  \includegraphics[width=.48\textwidth]{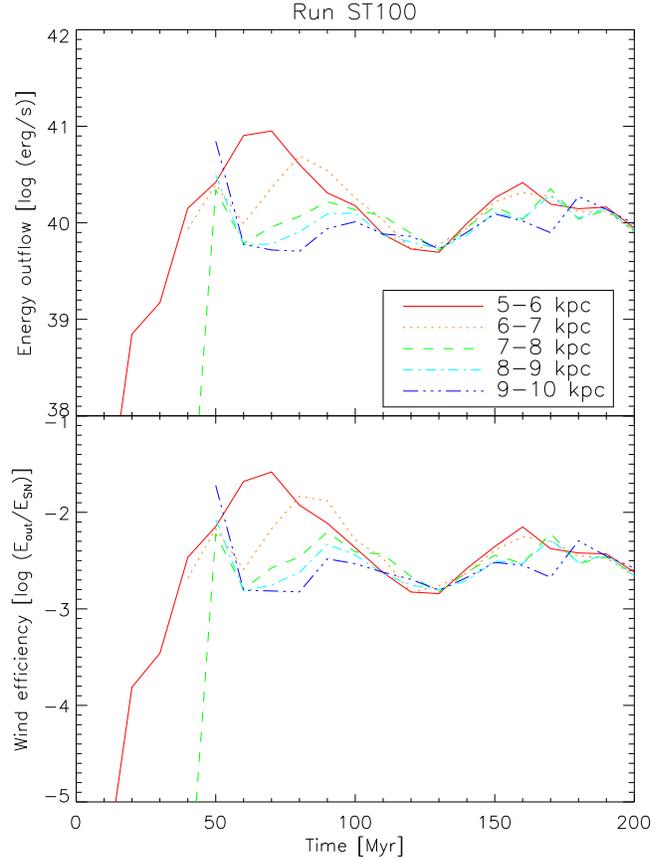}\\
  \vspace*{0pt}
  \caption{Analysis of simulation ST100. $Top:$ Energy flux rates through different shells of respective thickness $1\,\rmn{kpc}$. $Bottom:$ Efficiency of energy conversion, defined as the ratio of thermal plus kinetic energy carried by the wind $E_{\rmn{out}}$ to bulk energy released by supernovae $E_{\rmn{SN}}$.}
  \label{nrgflux_ST100}
\end{figure}

To obtain the net energy flux, we assume the same shells as before. 
\rev{The energy flux comprises kinetic and thermal components:}
\begin{eqnarray}
l_{E}(r,t)&=&k_{\phi}\frac{1}{\Delta r}\int_{0.04\pi}^{0.92\pi}\int_{-0.04\pi}^{0.04\pi}\int_{r_{\rmn{i}}}^{r_{\rmn{o}}} \nonumber \\ && \left(\frac{\rho(r,\theta,\phi,t)\,v(r,\theta,\phi,t)^2}{2}+\frac{p(r,\theta,\phi,t)}{\gamma-1}\right) \nonumber \\ && \cdot v_r(r,\theta,\phi,t)\,\rmn{d}\theta\,\rmn{d}\phi\,r^2\,\rmn{sin}\theta\,\rmn{d}r.
  \label{nrgflux}
\end{eqnarray}
The mean value for the net energy flux is averaged in the same way as the net mass flux, namely $L_{E}=\left<l_{E}(r,t)\right>$. \rev{Here, the polar contribution is a bit higher and we show in the appendix that we underestimate the energy fluxes by about 10~per~cent due to the polar cutout.} Again, the energy flux rates displayed in Figure~\ref{nrgflux_ST100} represent different shells of $1\,\rmn{kpc}$ thickness each, for different shell positions.\\ 
\begin{figure*}
  \centering
  \includegraphics[width=.93\textwidth]{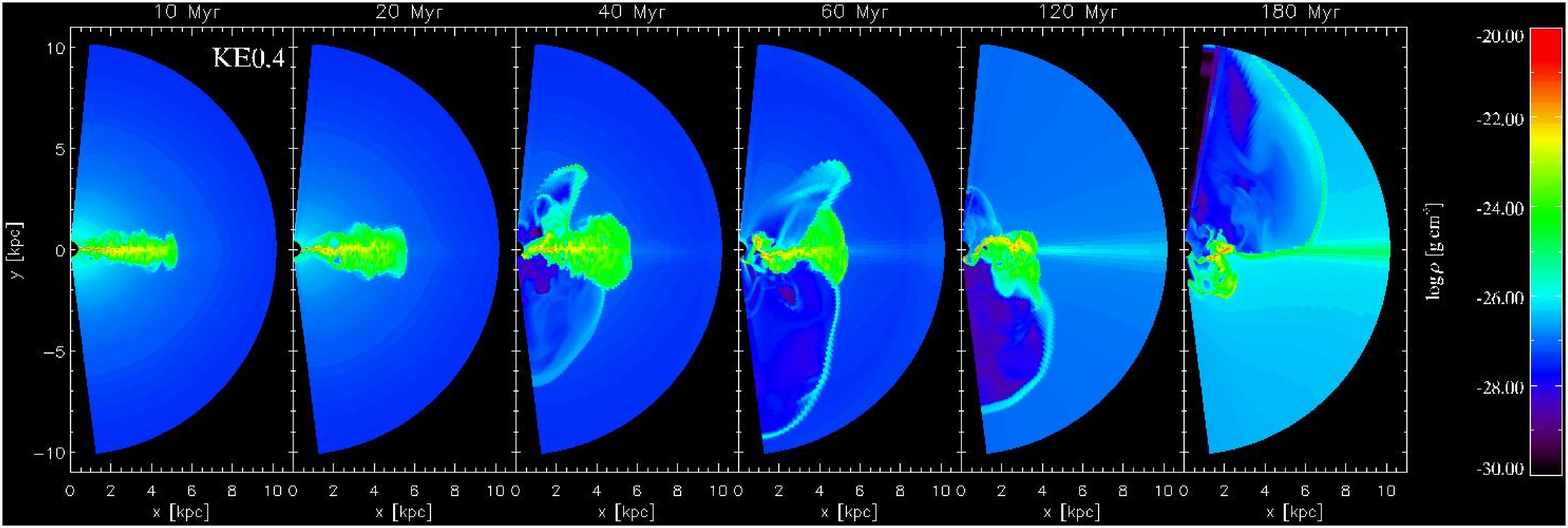}\\
  \includegraphics[width=.93\textwidth]{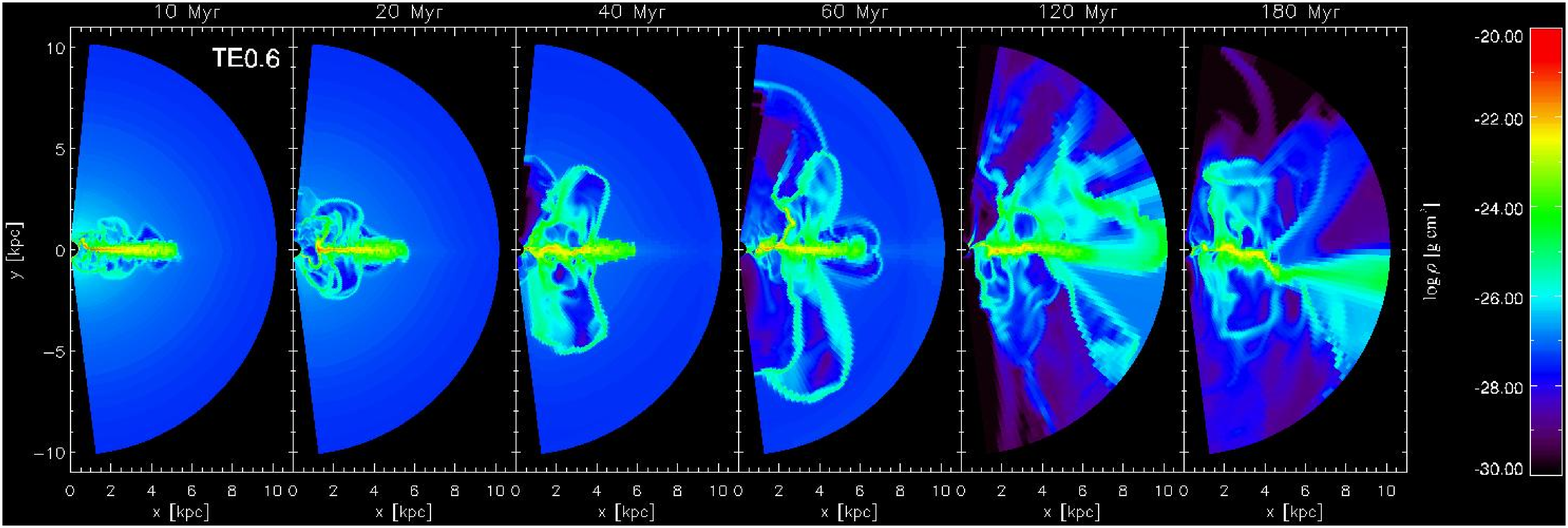}\\
  \includegraphics[width=.93\textwidth]{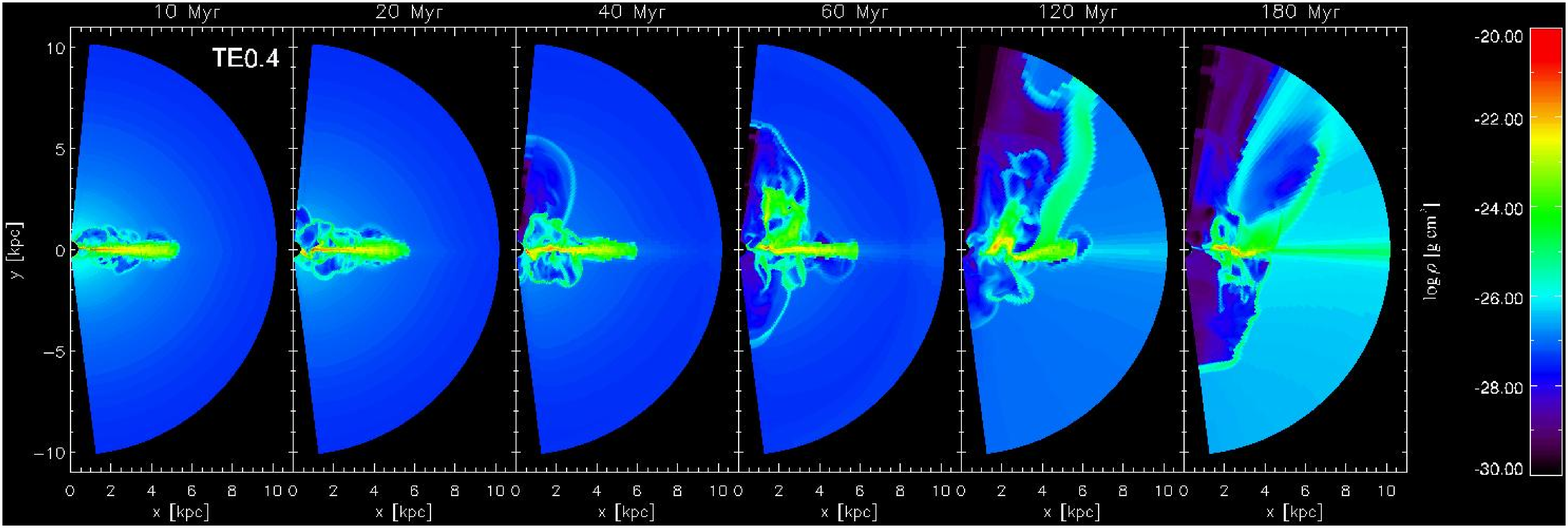}\\
  \vspace*{0pt}
  \caption{$Top:$ Simulation KE0.4 with each SN releasing $4.0\times10^{50}\,\rmn{erg}$ as kinetic energy only. $Middle:$ Simulation TE0.6 with each SN releasing $6.0\times10^{50}\,\rmn{erg}$ as thermal energy only. $Bottom:$ Simulation TE0.4 with each SN releasing $4.0\times10^{50}\,\rmn{erg}$ as thermal energy only. Snapshot times are identical to Figure~\ref{lgd_ST100}.}
  \label{lgd_KE-TE}
\end{figure*}
 Comparing the respective shells of measurement in Figs.~\ref{massflux_ST100} and ~\ref{nrgflux_ST100}, we can clearly see a convergence of the graphs with increasing shell radius. Measurements closer than $7\,\rmn{kpc}$ exhibit more pronounced extrema, and, in case of strong turbulent feedback or irregularities in the disc, may be prone to notable perturbations arising from the disc. If too close to the box boundary at $10.2\,\rmn{kpc}$, interactions with the boundary itself might distort the actual result in a few cases. Therefore, we choose the range in between $8\,\rmn{kpc} < r < 9\,\rmn{kpc}$ as the most reliable one.\\
All plots exhibit one more or less strong peak, which is the first shock front clearing the path for the wind yet to come. Any further peaks are a result of local and temporal concentrations of SN events; yet these anomalies will be mitigated as the energy outflow will stabilise over time. The basic level of energy carried by the wind is several $10^{47}\,\rmn{erg}\,\rmn{s}^{-1}$. So, with an average input of some $10^{50}\,\rmn{erg}\,\rmn{s}^{-1}$ in our models, we can define a wind efficiency as the ratio of wind energy to injected energy. The latter is stable on a level around $10^{-2.5}$, as is shown in the lower panel in Figure~\ref{nrgflux_ST100}.

\subsection{Kinetic energy-driven outflows}\label{sec:kwinds}

\begin{figure}
  \centering
  \includegraphics[width=.48\textwidth]{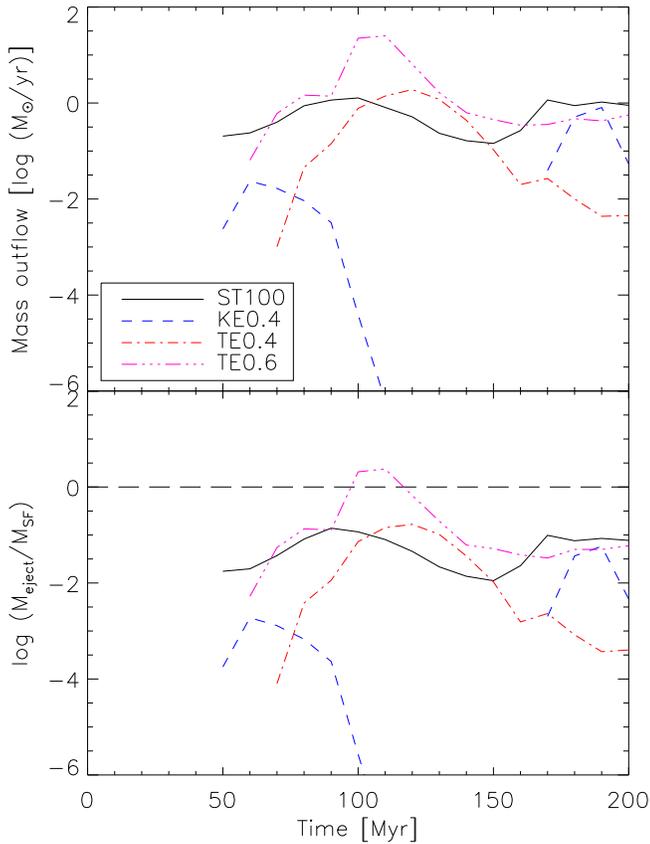}
  \vspace*{0pt}
  \caption{Absolute (top) and relative (bottom) mass flux rates, for different types and quantities of feedback energy. The solid black line represents model ST100 with a normal Sedov-Taylor energy distribution for comparison. The dashed blue, dash-dotted red and triple-dot-dashed purple line are the models KE0.4, TE0.4 and TE0.6, respectively.
  \label{massflux_KE}}
\end{figure}

\begin{figure}
  \centering
  \includegraphics[width=.48\textwidth]{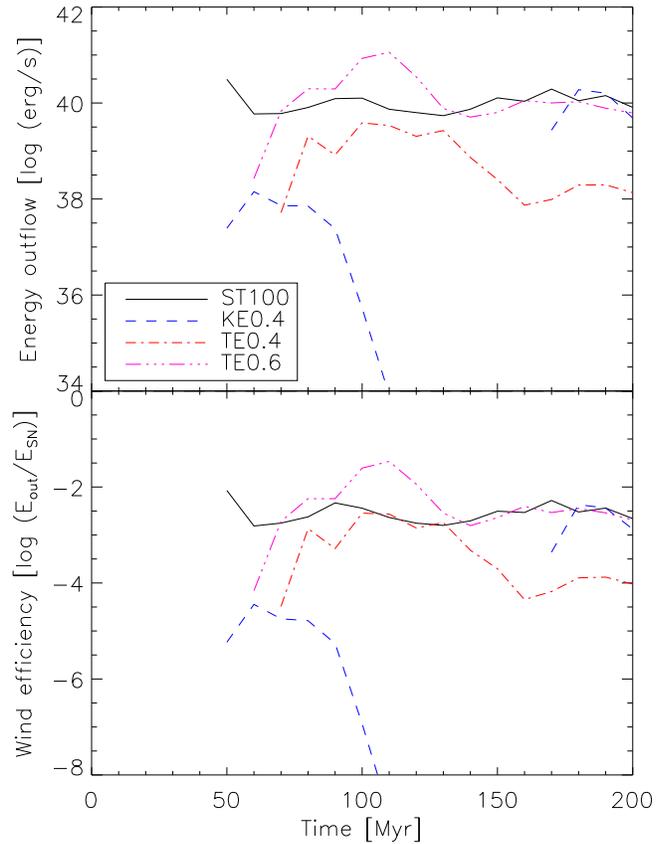}
  \vspace*{0pt}
  \caption{Absolute (top) and relative (bottom) energy flux rates, for different types and quantities of feedback energy. The line-styles and colours are the same as in Figure~\ref{massflux_KE}.}
  \label{nrgflux_KE}
\end{figure}

In order to compare directly the respective importance of the thermal and kinetic forms of energy injection, we have performed three simulations, where we inject only thermal energy or only kinetic energy (Figure~\ref{lgd_KE-TE}). Note that these simulations permit cooling in the halo, which subsequently reduces the environment pressure the wind has to overcome. The cooling halo is particularly necessary for the sake of the comparison in this section; without it an outflow may not be strong enough to leave the disc at all in some of the presented cases. In run TE0.6, we inject the thermal energy component, only, using the standard fraction of $0.6 \times 10^{51}\,\rmn{erg}$ per injected SN. This run has a slightly slower wind start, but later on is statistically indistinguishable from run ST100 regarding mass and energy outflow rates (Figures~\ref{massflux_KE} and ~\ref{nrgflux_KE}). Using only the 40 per cent kinetic energy (KE0.4), the outflow is much weaker: It has now a much harder time to get out of the disc. The part in the hemisphere with negative $z$ values is even dragged back by the ram pressure of the infalling halo (120 Myr). The outflow stalls completely between 110 and 120 Myr (compare Figs.~\ref{massflux_KE} and ~\ref{nrgflux_KE}). These results seem to indicate that the thermal energy part is the more important one for wind driving. We have also performed a run (TE0.4) with the thermal energy injection being reduced to the level of KE0.4. Here, the wind is also noticeably weaker, and the downwards going bubble also comes back. The statistics indicate a stronger outflow for TE0.4. However, the system is evidently just around the threshold, where it can drive a wind at all. Therefore, small changes might affect the result strongly. Remembering that our numerical scheme conserves the thermal energy better than the kinetic one (compare
section~\ref{blastwave}), we conclude that the differences between KE0.4 and TE0.4 are not significant.

\subsection{Bubble size}\label{sec:bubsize}
The last set of simulations presented in this study features a variation of the event size $\zeta_0$ introduced in Section~\ref{sec:fb}, \rev{above}. The event size specifies the number of SNe comprised in one single bubble. On average, for $100\,M_{\odot}$ of newly formed stars we expect one SN and a gas mass return through stellar winds and SN ejecta of $25\,M_{\odot}$. This in turn requires a minimum available mass of $7.5\times10^3\,M_{\odot}$ per cell for $\zeta_0=100$. However, there is a chance for a mass deficit to occur, typically in the outmost parts of the disc where the defined minimum density of $10^{-24}\,\rmn{g}\,\rmn{cm}^3$ is just reached, or in cells close to the inner radial boundary which exhibit small absolute angular diameters. On the other hand, the average cell mass will be $4.4\times10^4\,M_{\odot}$, which is well above the requirement for a 200-SN event. The mass deficit is not a severe issue, since in reality, the mass would come from neighbouring cells, and because the global error on the mass budget is small, no significant effect on the dynamics is expected. Locally, one might expect that we might artificially somewhat damp the kinematics in the large bubble simulations because of the slightly higher inertia in these runs. Yet, as we show below, we find that large bubble simulations exhibit the strongest winds.\\
Note that in some of the following simulations (NC20, NC50, NC100 and NC200) the threshold above which we inhibit radiative cooling is reduced below the halo equilibrium temperature of 600,000 $\rmn{K}$. We include these non-cooling simulations in addition to the ones with cooling at solar metallicity, in order to investigate possible effects of metallicity: For metal poor gas haloes, the cooling time is prolonged. Such galaxies will therefore likely have a hydrostatic halo as we describe it. For increasing metallicity, the thermal pressure will drop due to cooling but at the same time ram pressure due to the inflowing gas will increase \citep[compare ][]{DT2008}. With the approximations of solar metallicity cooling (ST) and non-cooling (NC) haloes, we try to capture the extreme cases, keeping in mind that a full parameter study in a Cosmological setup is clearly beyond the scope of this work. The values chosen for $\zeta_0$ in these simulations are 20, 50, 100 and 200 SNe, respectively (compare Table \ref{tab-sims}). In the following, the total SFR, the onset of the wind and its temporal development will be of particular interest. We will further investigate the mass and energy efficiencies in the same manner as above. It may seem reasonable to assume that, since smaller bubbles are situated much closer to each other than large ones, dense material in between will be further compressed until star formation sets in, thus providing a positive feedback to the SFR. Yet, large bubbles may proof more powerful when it comes to triggering the wind, and thus we could find that a larger $\zeta_0$, though providing little less energy input, results in a slightly more efficient wind.

\begin{figure}
  \centering
  \includegraphics[width=.48\textwidth]{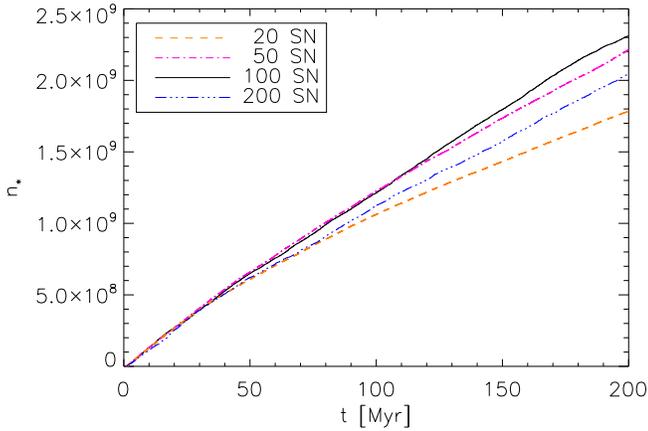}
  \vspace*{0pt}
  \caption{Total number of stars formed $n_{\star}$ for different respective amounts of SNe per bubble $\zeta_0$, indicated in the inset legend.}
  \label{SFR_bubbles}
\end{figure}

\subsubsection{Star formation}\label{sec:sf}

\begin{figure}
  \centering
  \includegraphics[width=.48\textwidth]{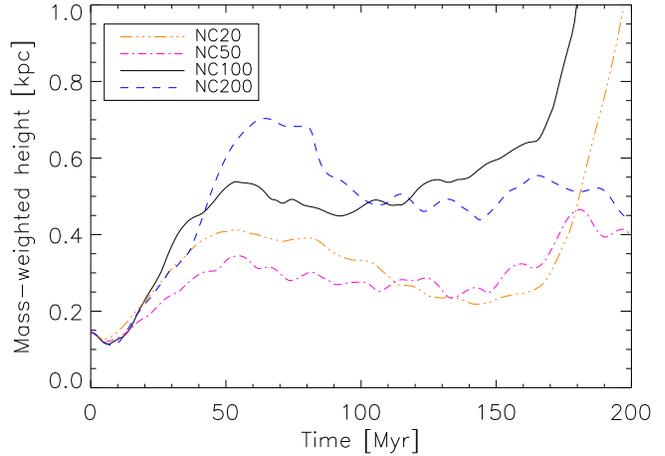}
  \vspace*{0pt}
  \caption{Mass-weighted height $h_M$ of gas above the disc midplane.}
  \label{massz}
\end{figure}

A look at Figure~\ref{SFR_bubbles} immediately reveals that the cumulative SFR for different bubble sizes undergoes little change within 8 per cent, until just before $50\,\rmn{Myr}$. This difference grows, being already around 24 per cent at $200\,\rmn{Myr}$. An explanation for this could be that large bubbles result in a violent blow-away of large gas portions, whereas small bubbles, due to their numerous occurrence, smear out the disc material over a comparatively large volume, reducing the chances for the gas to pile up in high amounts on any single spot. Both effects can result in a visible reduction of star formation, and hence the optimum range for star formation comes to lie in between 50 and 100 SNe per event.\\
In Figure~\ref{massz} we plotted the mass-weighted height $h_M$ of the gas above the disc midplane, which calculates as
\begin{equation}
  h_M=\frac{\int\,r\,\left|\rmn{cos}\theta\right|\,\rmn{d}m}{\int\,\rmn{d}m}.
\end{equation}

\begin{figure}
  \centering
  \includegraphics[width=.48\textwidth]{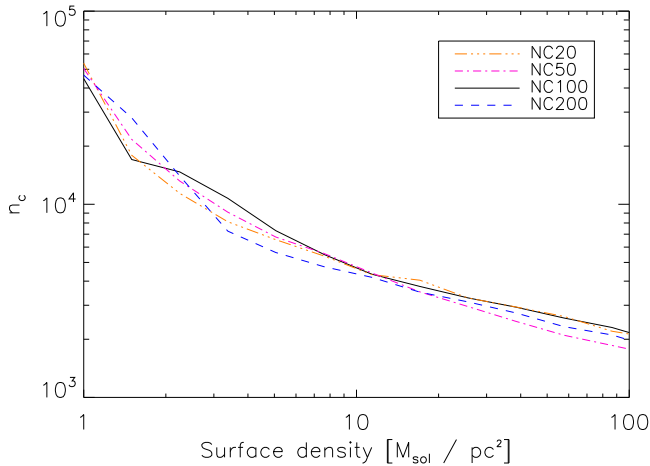}
  \vspace*{0pt}
  \caption{Surface density histograms for simulations NC20 - NC200, as indicated in the legend. We only take into account the region at radii $r<5\,\rmn{kpc}$ and sum up the columns of all the 200 snapshots of each simulation over the entire simulation time of $200\,\rmn{Myr}$.}
  \label{surfdens}
\end{figure}

The resulting value for $h_M$ indicates the average height of all gas portions in $\rmn{kpc}$ above the disc plane at any given time. We find that for NC200 $h_M$ is significantly larger \rvtwo{than} for NC100, but only between $40$ and $100\,\rmn{Myr}$, while NC20 and NC50 show comparatively little difference. NC100 and NC20 however increase strongly in the last $30\,\rmn{Myr}$. Increasing values mean that during this time much of the gas is torn out of the disc forming filaments, which constitute large quantities of gas unavailable for star formation. But if this were to be the reason for the lower SFR in NC200, we would expect the NC200 graph to dominate clearly from about $50\,\rmn{Myr}$ onwards. This possibility can hence be excluded.\\
In contrast, small bubbles of 20 SNe should have a smoothing effect on the overall density profile of the disc. The number of columns with respect to their density is visualised in Figure~\ref{surfdens}, whereas the total column number $n_{\rmn{c}}$ includes all columns within $r<5\,\rmn{kpc}$ and is integrated over the total simulated time span of $200\,\rmn{Myr}$. The curve for NC20 should exhibit more moderate values than its NC200 counterpart, whereas extreme values below $10\,M_{\odot}\,\rmn{pc}^{-2}$ and above about $60\,M_{\odot}\,\rmn{pc}^{-2}$ should be less present in the former. Columns of high density contribute most of all to the global SFR, and should be most present in the NC50 and NC100 curves. We find however, that neither of the four curves matches any of the expectations. Therefore, we can also exclude smoothing effects inside the disc from large numbers of small bubbles to be of notable effect to the SFR. This means that the SFRs in our simulations are set by a more complex interplay of processes.

\subsubsection{Mass and energy flux}\label{sec:fluxes}

\begin{figure}
  \centering
  \includegraphics[width=.48\textwidth]{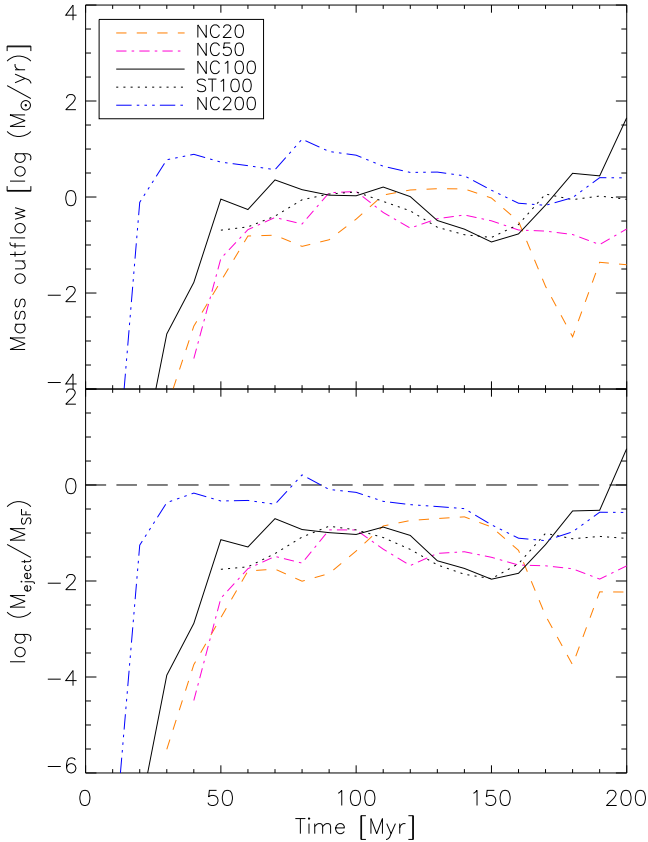}
  \vspace*{0pt}
  \caption{Mass flux rates in absolute ($upper$) and relative ($lower$) numbers for all four NC runs and ST100 (black dotted line). The relative values are normalised to the total mass of stars formed within the respective time.}
  \label{massflux_b}
\end{figure}

\begin{figure}
  \centering
  \includegraphics[width=.48\textwidth]{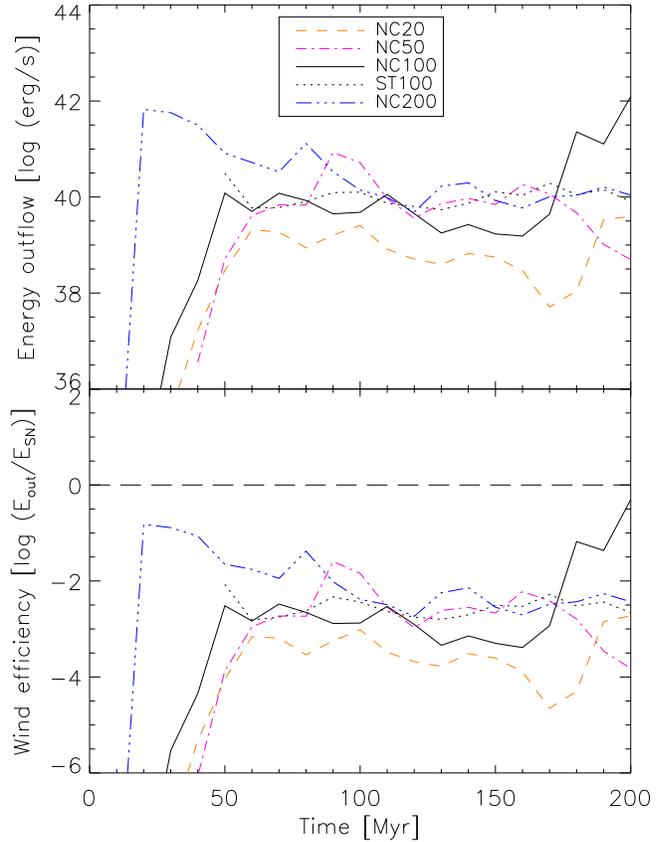}
  \vspace*{0pt}
  \caption{Energy flux rates in absolute ($upper$) and relative ($lower$) numbers for all four NC runs and ST100 (black dotted line). The relative values are normalised to the total energy release from SNe within the respective time.}
  \label{nrgflux_b}
\end{figure}

Figs.~\ref{massflux_b} and ~\ref{nrgflux_b} show the absolute and relative mass flux, and the absolute and relative energy flux, respectively, for the different bubble sizes. It has to be borne in mind, that after $100\,\rmn{Myr}$ the differences in the SFR become somewhat stronger (compare section~\ref{sec:acc} below). There is no doubt that the mass flow curve for NC200 starts earliest, and much higher than the others. Early starting curves are a clear indicator that the wind developed fast; in the case of NC200 it takes $20\,\rmn{Myr}$ for the wind to reach the radius of measurement at $8\,\rmn{kpc}$, giving it an average speed of nearly $400\,\rmn{km}\,\rmn{s}^{-1}$. Curves starting late suggest that the wind is setting in at a later point in time, but could also indicate a slower wind. The former case however applies to our simulations. The wind in NC100 starts early and still carries comparatively large mass. Of the two remaining ones, NC50 exhibits a stronger wind at a late start, whereas NC20 starts with little mass at an earlier time. When looking at Figure~\ref{nrgflux_b}, it becomes more obvious, that large SN bubbles show a tendency to start blowing a wind in a powerful way. The NC200 and ST100 energy curves stay roughly constant in time, whereas NC20 exhibits a more chaotic behaviour after the onset of the wind. While the order is not strictly maintained throughout the simulation time, there is a clear general trend for larger bubbles to produce higher mass outflow rates. This is also generally confirmed from the cumulative numbers (Table \ref{tab-flux}): The two \rev{large-superbubble} simulations have a mass outflow rate which exceeds the one of the two \rev{small-superbubble} simulations by about an order of magnitude. Run NC200 has formed 11 per cent less stars than run NC100, and still ejects 34 per cent more mass. Only for run NC 50, we find a 21 per cent smaller outflow rate in comparison to NC20, while the star formation rate is 24 per cent higher.\\
The trend is even more evident in the cumulative energy outflow rate (also in Table \ref{tab-flux}): For all the NC simulations, they increase monotonically with superbubble size, even if normalised to the star formation rate.

\subsubsection{Halo pressure}\label{sec:halopress}


\begin{table}
  \centering
  \begin{minipage}{75mm}
    \caption{Cumulative mass and energy flux values after $200\,\rmn{Myr}$ simulation time. The four bottom lines show the values for cooling halo models (ST) relative to non-cooling halo models (NC).}
    \begin{tabular}{@{}lcr@{}}
      \hline
      Run & Cumulative mass flux & Cumulative energy flux \\
      \hline
      ST20 & $2.7\times10^9\,M_{\odot}$ & $9.3\times10^{56}\,\rmn{erg}$ \\
      ST50 & $3.5\times10^8\,M_{\odot}$ & $1.1\times10^{57}\,\rmn{erg}$ \\
      ST100 & $1.0\times10^8\,M_{\odot}$ & $5.7\times10^{55}\,\rmn{erg}$ \\
      ST200 & $1.4\times10^9\,M_{\odot}$ & $3.5\times10^{56}\,\rmn{erg}$ \\
      NC20 & $7.7\times10^7\,M_{\odot}$ & $6.3\times10^{54}\,\rmn{erg}$ \\
      NC50 & $6.1\times10^7\,M_{\odot}$ & $7.2\times10^{55}\,\rmn{erg}$ \\
      NC100 & $6.2\times10^8\,M_{\odot}$ & $5.2\times10^{56}\,\rmn{erg}$ \\
      NC200 & $ 8.3\times10^8\,M_{\odot}$ & $6.3\times10^{56}\,\rmn{erg}$ \\
      ST20/NC20 & 35.1 & 147.6 \\
      ST50/NC50 & 5.7 & 15.3 \\
      ST100/NC100 & 0.16 & 0.11 \\
      ST200/NC200 & 1.7 & 0.56 \\
      \hline
      \label{tab-flux}
    \end{tabular}
  \end{minipage}
\end{table}

In Table \ref{tab-flux} the complete set of runs ST20, ST50, ST100 and ST200 is compared to their respective NC counterparts. The displayed values are the cumulative mass and energy flux rates until $200\,\rmn{Myr}$, in absolute numbers, $\int_0^{200\,\rmn{Myr}}L_M(t)\,\rmn{d}t$ and $\int_0^{200\,\rmn{Myr}}L_E(t)\,\rmn{d}t$, respectively. For each bubble size, the flux value of the respective ST run is normalised by the value for the respective NC run.
It is obvious that for the smaller bubble sizes, $\zeta_0=20$ and $\zeta_0=50$, the outflow is stronger in the absence of thermal halo pressure. Moreover, ST20 and ST50 feature one major outburst each, where a massive local concentration of feedback energy leads to the ejection of a large share of hot gas from the disc. However, if $\zeta_0=200$, a steadily blowing wind arises also for the thermally pressurised halo; we find both mass and energy outflow rates for ST200 and NC200 to range in the same order of magnitude, respectively. $\zeta_0=100$ represents a special case, where an exceptionally large filament is torn out of the disc after $170\,\rmn{Myr}$, which accounts for the bulk of mass and energy (also compare Figure~\ref{STtoNC}). If this phenomenon is neglected, the flux values for ST100 and NC 100 would be of comparable magnitude.

Figure~\ref{STtoNC} shows one snapshot from all eight runs at the same time of $200\,\rmn{Myr}$. We find for $\zeta_0=20$ that in both cases the small bubble size only triggers a weak wind. In NC20, filaments bordering the upper and lower wind conus are absent, indicating that the halo pressure has already begun to force the wind conus back into the disc. In ST20 we find the wind to be asymmetric, being at least stable on one side of the disc. The same applies to ST50, where the wind is also dominant on one disc side only. NC50 in contrast developed a biconically stable wind, however, the conus is already in the process of being crushed. The wind in NC100 has ceased entirely; instead we can see the disc being just a few Myr before complete disruption - which also explains the enormous mass and energy outflow rates towards the end of NC100. ST100 on the other hand exhibits a clear biconical wind structure, with a wind steadily blowing in both directions. Stable winds also occur in NC200 and ST200. This supports our assumption that large superbubbles generally seem to boost the overall strength and steadiness of the wind. Furthermore it appears that for smaller bubbles the environment pressure becomes important: If the halo is thermally pressurised, winds arise but cannot overcome the halo pressure in the long term. In case of a cool, less pressurised halo, winds are on the brink of developing towards a stable, steady state; asymmetric developments with at least one of two coni being stable are not unlikely.\\

In summary, mass and energy outflow rates in the NC runs consistently show the same trend: if the event size $\zeta_0$ is varied, the outflow rates for 
\rvtwo{large $\zeta_0$ will} tend to start comparatively high, and change barely over time. \rvtwo{Small $\zeta_0$ will} cause the wind to set in less forcefully, and, as is the case with NC20, undergo occasional drops in strength. The efficiency of mass ejection in our NC models \rvtwo{is typically} between $\sim10^{-2}$ and unity. The efficiency of feedback energy conversion exhibits a convergence for most runs against $\sim10^{-3}$, while values of $\sim10^{-2}$ are still common, and  $\sim10^{-1}$ is already rare.\\
For the ST runs \rvtwo{(low halo pressure)}, no clear trend can be discerned. A high halo pressure efficiently pushes smaller bubbles back into the disc, but low halo pressure enables bubbles of all event sizes to enter the halo overpressured and keep expanding. Therefore, the outflow properties do not depend systematically on the bubble size in the latter case. Instead, they tend to be dominated by single events, like the high concentration of SN bubbles leading to a violent ejection of large gas masses in ST20 and ST50.

\section{Discussion}\label{discu}

\subsection{Methodical accuracy}\label{sec:acc}
We have produced successful models of LBGs launching galactic outflows, in order to shed some light onto the exact mechanisms responsible for the onset of a galactic wind. These models feature a realistic galaxy setup with superbubble events, similar to the setup invoked by \citet{DT2008}. Our equilibrium setup allows us to investigate the reaction of the system to systematic changes of parameters like the halo pressure or the superbubble size. Our methods described in sect. 3 comprise the most important physics, however, some simplifications had to be made which require further discussion.

Firstly, our SN bubbles were triggered randomly, in accordance to the Kennicutt-Schmidt law for star formation (\ref{snprob}), which is correlated to the column density $\Sigma$, but neglects the volume density of the constituent cells within the column. One might argue that it would be better to let the SN probability increase with the density. This in turn will mean that more bubbles occur deeper within the disc and will thus have a harder time reaching the halo. The net effect would be an overall mitigation of the wind by an unknown factor. In consequence, energy would be converted even less \rvtwo{efficiently}.

We further had to implement a lower volume density threshold for cells to count as part of the disc and to amount to the surface density of their specific column. Note that the value used for our models, $2\times10^{-24}\,\rmn{g}\,\rmn{cm}^{-3}$, is just a crude estimate for the lowest density regions found in the $10^4\,\rmn{K}$ gas phase of the ISM, and thus allows for some variation. For instance, a lower threshold will open up a regime of rarefied cells surrounding the disc as is currently defined. This will have an effect on the distribution of the SN events, allowing for a bubble to blow out into the halo with less resistance. Though, the change in total will likely be of little effect regarding the wind strength - note that such rarefied cells will likely contain just around $100\,M_{\odot}$. This would definitely call for the modification of our probability function, which, if applied, would make an event in these cells extremely unlikely.
\begin{figure*}
  \centering
  \includegraphics[width=.93\textwidth]{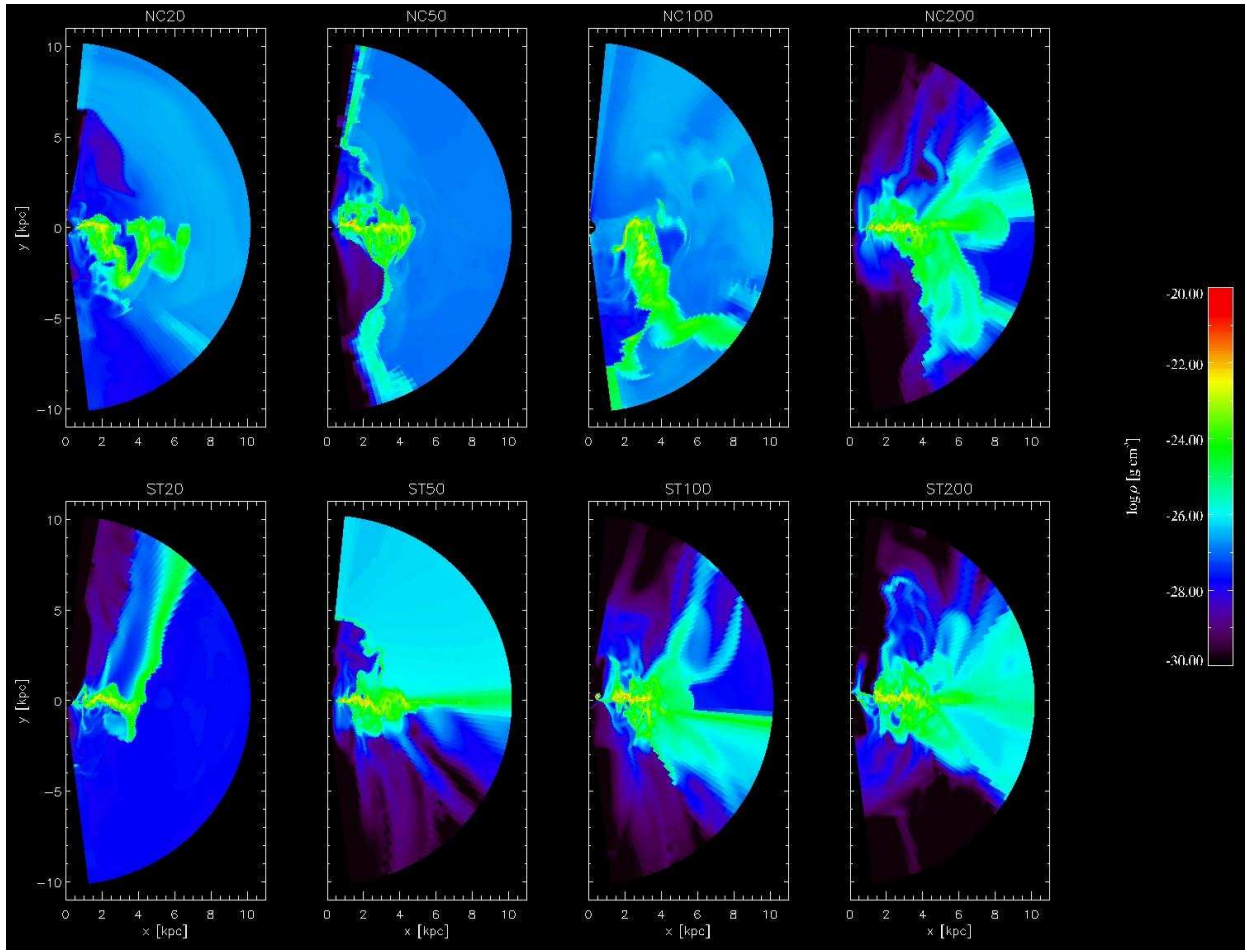}
 \vspace*{0pt}
  \caption{$Top$ $row:$ Simulation set NC at $t=200\,\rmn{Myr}$ for $\zeta_0=$20, 50, 100 and 200, from left to right respectively. $Bottom$ $row:$ Simulation set ST at $t=200\,\rmn{Myr}$ for the corresponding values of $\zeta_0$.}
  \label{STtoNC}
\end{figure*}

The star formation in our models is determined by a local Kennicutt-Schmidt law, and converges with increasing resolution. It is also sensitive to local events, such as material ejections and the bubble size. The latter is clearly a significant effect: In our resolution study (Figure~\ref{SFRres}), we find that the number of stars formed after $200\,\rmn{Myr}$ agrees within 26 per cent. However, there is a convergence for resolutions finer than 36 pc. If we disregard the 65 pc resolution, the deviation already shrinks to nine per cent. Varying the bubble size yields a change in star formation of about 24 per cent. Larger bubble size leads to stronger star formation, yet very large bubbles seem to lead to such a strong outflow that the star formation gets weaker again. We believe that this feature of the model is realistic.

We have also investigated the dependence of the outflow rates on resolution. Apart from the 65 pc case (R65), which also shows a stronger deviation in the star formation rate, there is no strong or systematic deviation among all the runs with better resolution. \rev{One might argue that the smaller event sizes lead to less well-resolved 
superbubbles, and thus the higher outflow rates for larger superbubbles might be explained by such a resolution effect. The effect is however not present for the ST runs. Because the pressure in the disc is the same for ST and NC runs, the initial bubble sizes should be similar, which make resolution effects an unlikely explanation for the trends with bubble size.}

\rev{Our simulations assume large-scale rotational symmetry. This means that the structure of the ISM is considerably restricted. Clumping in the azimuthal direction or spiral structure may increase the local star formation density, while leaving the average value constant. In a gas rich galaxy such as considered here local feedback effects should then also become important \citep[e.g.][]{Hopkea12}. In general, one should expect that superbubbles would break out of the dense gas more easily, as the effective surface of the star-forming regions is increased. This should reduce the critical bubble size for winds to occur and might increase the outflow rates. 
For these reasons our results should be regarded as qualitative, only.}

\subsection{Wind drivers}\label{sec:wdrive}
\rev{In our simulations, winds form from sequences of superbubbles: Individual superbubbles 
break out of the disc and remain in the lower halo for about $10^7$~yrs. We find hardly any bubbles to be affected by buoyancy. Either bubbles simply collapse after having overexpanded and fall back to the disc, or they are 
fed by subsequent superbubbles which eventually drives a larger scale outflow. For our standard run, this happens above a critical star formation density of 0.1~$M_\odot$~yr$^{-1}$~kpc$^{-2}$. We show that the effect as such may
happen even if we inject either only the thermal energy or only the kinetic energy component. The outflow is then however attenuated.}
\rem{Our work concentrates on the mechanisms behind galactic outflows. We have shown that buoyancy can drive bubbles out of the disc, but by itself is not powerful enough to trigger fast-flowing galactic winds without any supplementary yield of energy, even not in our present simulations, which if anything overestimate buoyancy. We find that the thermal energy component by itself is sufficient to drive a wind. Whether or not we add the kinetic energy does not significantly change the result. We find a much weaker outflow if we use the kinetic energy component, only. Even in simulations with the
same overall energy injection, but once thermally and once kinetically injected, we find that the thermally driven wind is much stronger.} There is a good reason why we should expect such a behaviour: A pressure supported bubble will simply expand into the direction of the strongest pressure decline, i.e. radially outward, once the halo is reached. Gas which just has its kinetic energy may not accelerate as efficiently by this pressure gradient. On the contrary the pressure force works the other way, because the bubble is under-pressured much faster. Moreover, we find the outflow energy to vary by a factor of 100 for different event sizes (Figure~\ref{nrgflux_b} and Table \ref{tab-flux}), while at the same time NC200 produces a strong wind, and NC100 a weak one, just strong enough to enter the thermally pressured halo. 
\rev{Hence, only superbubbles comprising at least around 100~SNe seem useful in order to generate a wind, slightly more if the halo is still relatively massive but cools inefficiently, as modelled in our NC runs.}
\rem{These two orders of magnitude in outflow energy make the difference between existence and absence of a wind. For the different feedback energy types however, the outflow energy already differs by $10^6$ (compare Figure~\ref{nrgflux_KE}), with the outflow for the KE0.4 run temporarily coming to a complete breakdown.
However, because thermal energy is conserved better by our numerics, we may underestimate the effects of the kinetic energy injection.}

\rem{It should be pointed out here, that additional kinetic energy inside a disc might generally be provided e.g. by turbulence resulting from filamentary inflow or from magneto-rotational instabilities. These, in addition to the SN feedback might make up for a total kinetic energy supply high enough to trigger significant outflow. However, if we are to deal with solely feedback-driven winds, excluding sources of energy not directly related to SN feedback, our results indicate that thermal energy drives the outflow in a more constant manner than kinetic energy, even if both are injected in equal quantities. Complete suppression of the outflow, as is the case with kinetic energy in between $110-170\,\rmn{Myr}$ does not occur with thermal energy-driving.\\
We have found for run KE0.4, that kinetic energy is converted highly inefficiently. Hence, we could argue that, no matter what the actually provided amount of kinetic energy is, the chances of resulting in a continuous outflow are generally small. Even stronger kinetic energy feedback would then, more likely, result in partial or even complete disruption of the gas disc as such. This becomes clear when we take a look at the morphology of our model discs. Our simulations already show heavy distortions of the disc by the injected feedback energy, suggesting that they are close to the kinetic energy level necessary for disruption. If then a comparable amount of additional kinetic energy is spread equally inside the disc, we expect exactly this to happen. Regarding the leading shock fronts in this case, if a shock front made of compressed gas is allowed to expand only by its momentum initially received, it will quickly slow down and collapse again due to over-expansion.\\
To ensure further growth, the shock front needs support by significant overpressure of the underlying rarefied region, which will be continuously provided by further superbubbles strong enough to penetrate the halo. The cool, dense phase of the ISM is torn out of the disc to form the wind's filaments, which feature prominently in optical emission line studies. Aside from that, this phase is required for star formation in the first place, making it the key ingredient to trigger the feedback cycle. It therefore seems reasonable to assume that the key to driving a wind lies within the existence of a multiphase ISM, featuring a cool ($<10000\,\rmn{K}$), dense phase as well as a hot ($>10^6\,\rmn{K}$), rarefied phase at high pressure. Comparing to observations of wind galaxies in general \citep[e.g. ][]{VCBh2005,SBh2010}, the steady thermally driven solutions which channel the energy efficiently into the wind and hence enable a comparatively calm disc also seem to be preferred.\\}

Furthermore, in sect. 4.4 we have seen that the bubble size matters during the phase where the wind is launched and breaching through the inner halo regions. In reality, bubbles will not be all of the same size but rather occur in a wide range from single, isolated SNe to a few hundred per bubble. Here, we show that the larger superbubbles matter the most for galactic winds. However, for more realistic event size distributions the mass and energy flux in the resulting wind might converge earlier and exhibit fewer and smaller peaks. For the time being, we will leave this matter open for future investigation.

The bubble size has turned out to be relevant firstly for the initial shock wave, and secondly in the steady wind phase. Larger bubbles give a more powerful rise to the wind, and will keep their strength at higher, roughly constant levels for a long time. In LBGs which presumably blow winds continuously at a steady level, the bubble size could be important. Moreover, during \rev{a} starburst phase, where the wind is often young and after which star formation will decrease rapidly, the bubble size might play a role.

The thermal halo pressure determines whether or not a stable wind phase develops in the first place. If the pressure is too high, the wind may very well be unable to proceed too far from the disc. How far it can go depends on the bubble size. Winds set up by small bubbles will stop early and in some cases collapse back onto the disc entirely, whereas winds resulting from large superbubbles have a good chance of escaping the halo no matter the halo pressure.

\rev{\subsection{Comparison to observations}
We model galaxies with masses, star formation rates, and star formation densities 
comparable to LBGs. We find that our superbubble mechanism starts to produce
outflows from star formation densities around  0.1 ~$M_\odot$~yr$^{-1}$~kpc$^{-2}$, which would predict outflows in essentially all LBGs, as observed.
While we do not follow the ionisation state of the gas, it is clear that the gas around 1~cm$^{-2}$ in our simulations is highly turbulent, often escaping the disc
(Figure~\ref{STtoNC}). This is in good agreement with the generally observed high level of turbulence in LBGs,
e.g. the metal line observations in \citet{Lawea12}.
The superbubble mechanism produces naturally weak, radiative shocks in the halo, which we observe directly in the beginning of the simulations (Figure~\ref{slices}).
\begin{figure}
  \centering
  \includegraphics[width=.48\textwidth]{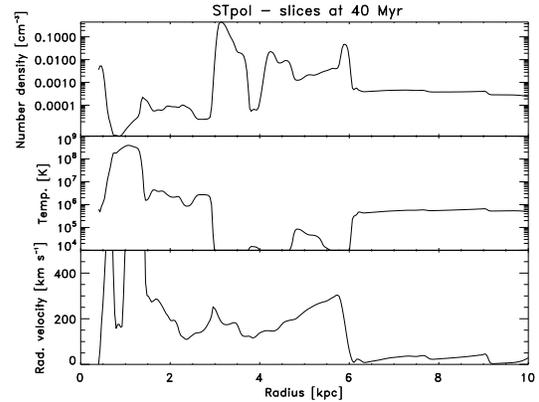}
  \vspace*{0pt}
  \caption{Radial slices of number density, temperature and radial velocity at $\theta=0.33\pi$ for run STpol at 40~Myr. Temperature values of 10$^4$~K correspond to our minimum temperature and are in reality likely lower further away from the disc.
The plot shows outflowing cold shells at velocities around 200~km~s$^{-1}$.
The one at 3 (4,6)~kpc has a column of about $10^{19}$~cm$^{-2}$ ($10^{18}$~cm$^{-2}$, $10^{18}$~cm$^{-2}$).
  \label{slices}}
\end{figure}
These shells would plausibly be observed as Ly~$\alpha$ absorption systems 
with column densities around $10^{18}$~cm$^{-2}$ to $10^{19}$~cm$^{-2}$, and velocities around 200~km~s$^{-1}$, which compares favourably to observations
\citep[e.g.][]{Verhea08}. If anything, one would like to increase the halo density
by a factor of a few, in order to cover the full range of observations.

At later times, these shells leave the computational domain. In many of our simulated galaxies, the shells actually come back and may start over. This happens when we 
moderate the feedback mechanism or superbubble size. 
Overall, the mechanism seems well able to explain wind shells in LBGs.}

In observations of nearby galactic winds, one frequently finds energy efficiencies of order ten per cent \citep{VCBh2005}. We find much less in our simulations. A similar discrepancy is seen in the mass outflow rates. It is well possible that these are different classes of objects. \rev{If the high wind efficiencies would also be confirmed for LBGs, it might point to some effects we might still be missing in our simulations.
In particular, if the galactic halo is cleared out in some regions, simulations which 
more accurately treat the feedback mechanisms in the disc and which find much higher mass loss rates from the disc might become more applicable.}


\section{Conclusions}\label{sec:conc}
We have performed hydrodynamic simulations 
\rvtwo{in spherical wedge geometry with polar cutouts}
with the grid-based 3D code NIRVANA, setting up a disc-halo system close to hydrodynamic equilibrium. Stars are formed in agreement to a local Kennicutt-Schmidt law, and 
\rev{superbubbles are injected according to a preset number of instantly occurring SNe. 
\rvtwo{Thus, the structure of the interstellar medium is simplified to large-scale effects. Spiral-arm structure is not taken into account.}
In regions of the galactic discs, where a critical star formation density is exceeded, series of superbubbles expand into the galactic halo at slightly supersonic speed,
lead by radiative shocks which may produce cold gas \rvtwo{shells.}

\rvtwo{We parametrise the outflow strength via mass and energy loss rates. 
For our chosen galaxy setup and feedback implementation, the outflow strength depends on the details of 
the feedback implementation, the stellar content per superbubble and the halo pressure:
We do not get significant outflows if we neglect the thermal energy part in the superbubble injection, and only retain the kinetic part.
Outflows are stronger for bigger individual superbubbles, if the halo pressure is not too large. For large halo pressure, outflows are suppressed
for superbubbles below a critical size. Our absolute outflow rates are uncertain due to the assumption
of uniform stellar content per superbubble in a given simulation, and the simplifications in the modelling of the structure of the interstellar medium
and the superbubble injection mechanism. }

The simulated galaxies are in overall good agreement with LBGs regarding
mass, star formation rates, star formation densities, gas kinematics and expanding shell characteristics. This suggests that the interaction between superbubble-driven winds and a heavy gaseous halo is a good candidate to explain the characteristics of LBGs.}

\rem{The feedback energy released this way eventually leads to a number of effects considered to play a role in the development of galactic winds.
We study four potentially important factors for wind driving, namely the thermal energy contribution, the kinetic energy contribution, buoyancy of SN bubbles and the local concentration of SNe which determines pressure and size of the superbubbles. It could be shown that the main launching mechanism is the thermal energy contribution and the amount of pressure it provides to a bubble. In consequence, we consider a multiphase ISM to be essential for the numerical simulation of galactic winds. The feedback of thermal energy alone is capable of converting up to about a factor of $10^{-2}$ times the available energy into the outflow. In comparison, kinetic energy can increase the strength of a wind but is very unlikely to be capable of launching a wind by itself. Buoyancy of superbubbles provides only around one per cent of the energy carried by the wind, which alone is insufficient to set up a wind as well. The wind strength grows with the superbubble size, and is significantly affected by the halo pressure. Overall, our wind strengths are rather on the low side, which might point to details of superbubble physics not yet captured by our simulations.}



\section*{Acknowledgements}
The authors would like to thank the CAST group for the regular meetings including several useful suggestions on key aspects of this paper. We also would like to thank the Excellence Cluster Universe and the DFG (KR 2857/3-1) for the financial support, which made this project possible. Our cordial thanks further goes to \rev{Barbara Ercolano, Mark Westmoquette, Anne Verhamme and Daniel Schaerer for helpful discussions.} We are also grateful to Ralf-J{\"u}rgen Dettmar for fruitful discussions on the topic during his visit to USM. Special thanks finally goes to Christian Tapken and Matthew Lehnert for helpful comments on our project during the DFG conference in Potsdam, September 2010. 
\rev{We also thank the anonymous referee for very useful comments that considerably improved this paper.}

\appendix

\section[]{Control runs}
\begin{figure}
  \centering
  \includegraphics[width=.46\textwidth]{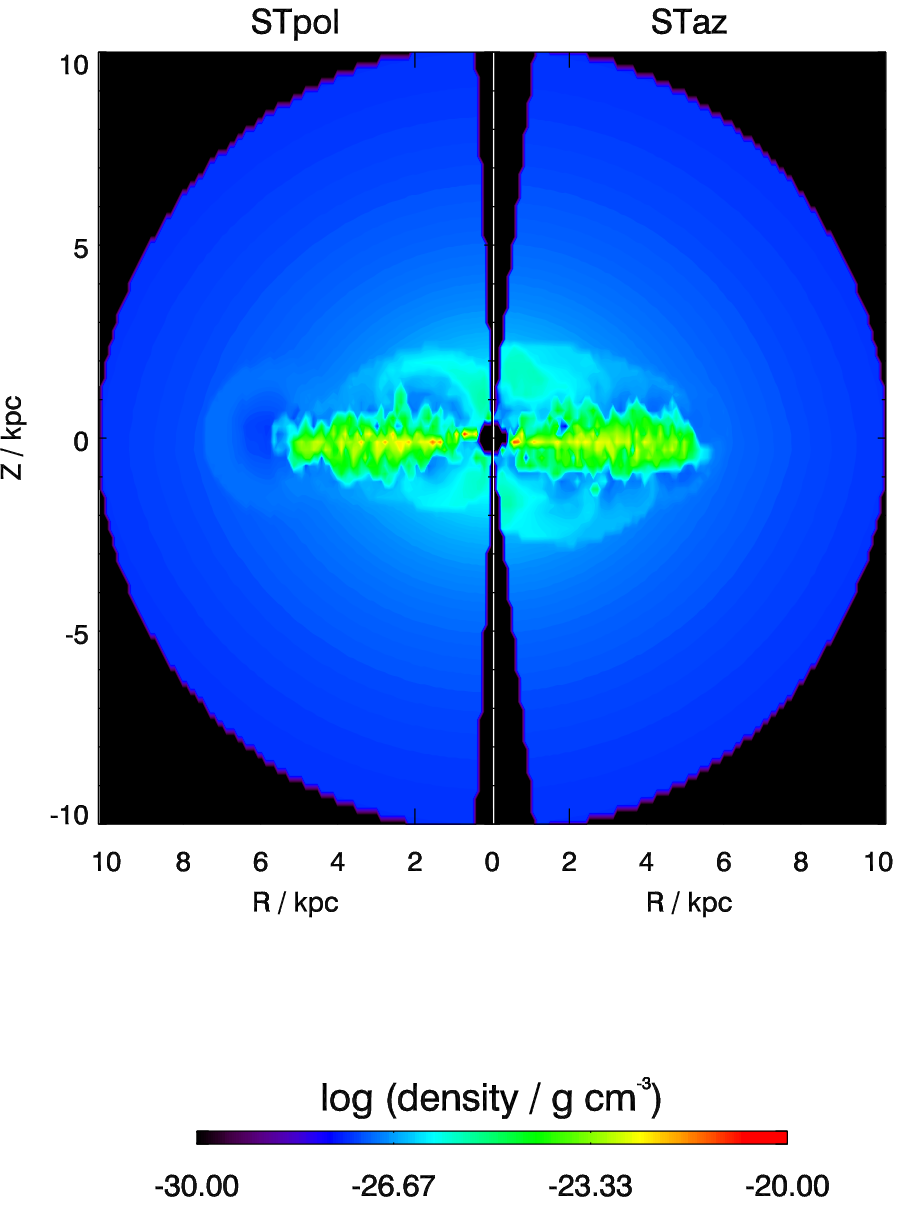}
  \vspace*{0pt}
  \caption{Comparison of runs STpol (left side) and STaz (right side) at 10~Myr. 
    The structure of the of the disc ISM and the expansion of the superbubbles in the inner part of the disc above and below the disc is very similar. These features are also similar to the basic run ST100 (Figure~\ref{lgd_ST100}).}
  \label{fig:ctrlplot}
\end{figure}

We have performed control runs to check for the impact of our choice of the size 
of the azimuthal wedge and the polar cutout. Run STpol is identical to ST100 apart 
from the polar cutout having been reduced from $0.04\pi$ to $0.02\pi$. Run STaz 
is again identical to run ST100, but the we simulate half the azimuthal angle, 
instead of just $0.08\pi$. 
We show 
a comparison of meridional density plots in Figure~\ref{fig:ctrlplot} at 10~Myr, 
when the wind first develops. From the plot, it is clear that the general structure 
of the disc ISM, and the extent of the superbubble expansion into the halo are 
very similar also to the corresponding plot for run ST100 (Figure~\ref{lgd_ST100}). 
It is beyond the scope of this article to investigate the effect of 
prominent non-axisymmetric structures on outflow rates, like e.g. spiral arms. 

\begin{figure}
  \centering
  \includegraphics[width=.46\textwidth]{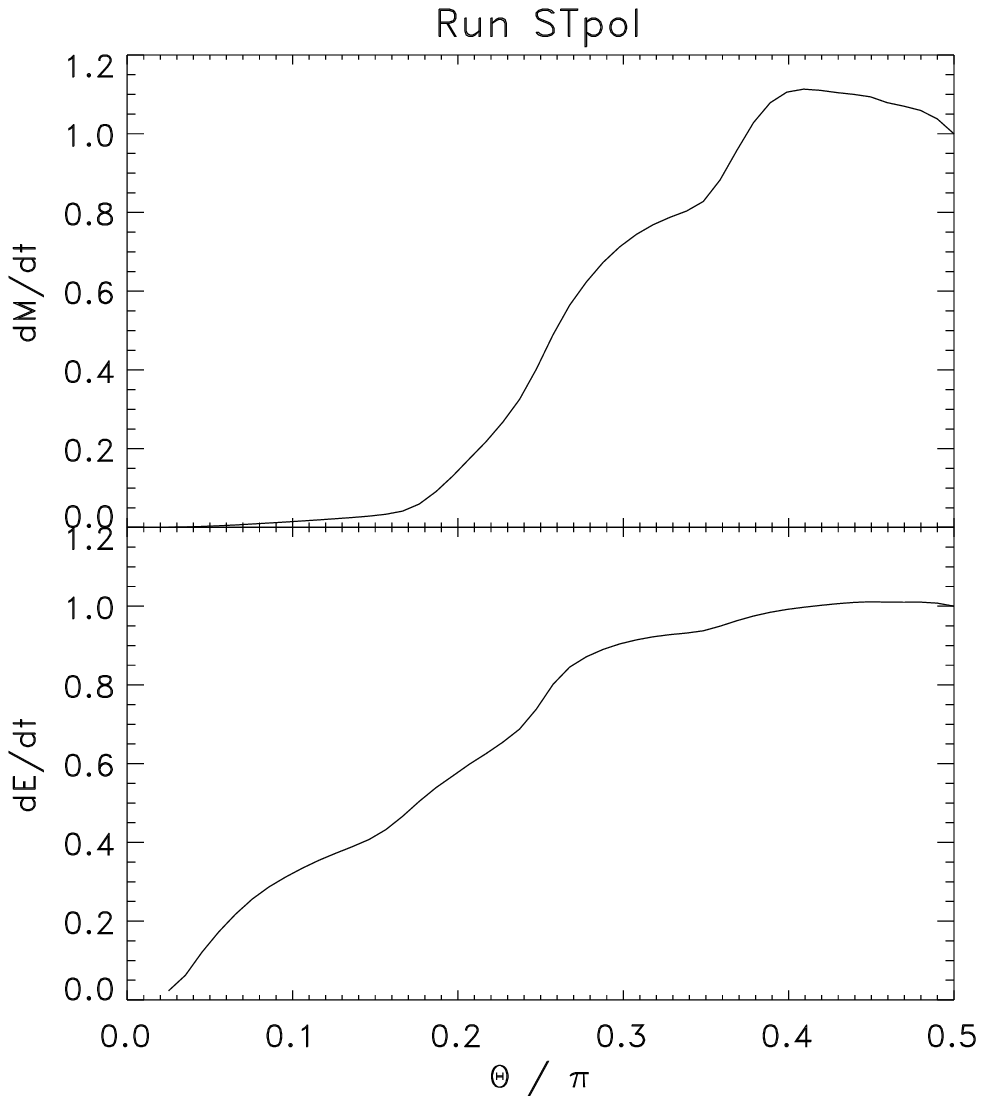}
  \vspace*{0pt}
  \caption{Cumulative mass flux (top) and energy flux (bottom) over solid angle for run STpol. As in the main paper, we average the fluxes over the shell between 8 and 9 kpc. While the mass flux in the polar cutout region is low anyway, the polar cutout needs to be small in order to capture most of the energy flux.}
  \label{fig:ctrlfluxes}
\end{figure}

The run of the outflow rates in run STpol are within a factor of a few
similar to the ones of ST100. This is within the general scatter of the ST runs, 
and thus expected.
 We show the cumulative mass and energy fluxes 
over the polar angle in Figure~\ref{fig:ctrlfluxes}, averaged over all the 201 snapshots 
of the simulation. 
In the standard runs, 0.8 per cent of the solid angle is cut out. In run STpol, only 
0.2 per cent of the solid angle is still cut out. The extra 0.6 per cent of solid angle 
contribute 1~percent of the mass flux and 10~per~cent of the energy flux. 
The differential energy flux declines towards the polar axis.
This should be regarded as the typical uncertainty due to the polar cutout.

\bsp

\label{lastpage}

\end{document}